# A non-singular theory of dislocations in anisotropic crystals

Giacomo Po[a], Markus Lazar[b], Nikhil Chandra Admal[c], Nasr Ghoniem[a]

[a]*Department of Mechanical and Aerospace Engineering, University of California Los Angeles, Los Angeles, CA 90095*
[b]*Department of Physics, Darmstadt University of Technology, Hochschulstr. 6, D-64289 Darmstadt, Germany*
[c]*Department of Materials Science and Engineering, University of California Los Angeles, Los Angeles, CA 90095*

**Abstract**

We develop a non-singular theory of three-dimensional dislocation loops in a particular version of Mindlin's anisotropic gradient elasticity with up to six length scale parameters. The theory is systematically developed as a generalization of the classical anisotropic theory in the framework of linearized incompatible elasticity. The non-singular version of all key equations of anisotropic dislocation theory are derived as line integrals, including the Burgers displacement equation with isolated solid angle, the Peach-Koehler stress equation, the Mura-Willis equation for the elastic distortion, and the Peach-Koehler force. The expression for the interaction energy between two dislocation loops as a double line integral is obtained directly, without the use of a stress function. It is shown that all the elastic fields are non-singular, and that they converge to their classical counterparts a few characteristic lengths away from the dislocation core. In practice, the non-singular fields can be obtained from the classical ones by replacing the classical (singular) anisotropic Green's tensor with the non-singular anisotropic Green's tensor derived by Lazar and Po (2015b). The elastic solution is valid for arbitrary anisotropic media. In addition to the classical anisotropic elastic constants, the non-singular Green's tensor depends on a second order symmetric tensor of length scale parameters modeling a weak non-locality, whose structure depends on the specific class of crystal symmetry. The anisotropic Helmholtz operator defined by such tensor admits a Green's function which is used as the spreading function for the Burgers vector density. As a consequence, the Burgers vector density spreads differently in different crystal structures. Two methods are proposed to determine the tensor of length scale parameters, based on independent atomistic calculations of classical and gradient elastic constants. The anisotropic non-singular theory is shown to be in good agreement with molecular statics without fitting parameters, and unlike its singular counterpart, the sign of stress components does not show reversal as the core is approached. Compared to the isotropic solution, the difference in the energy density per unit length between edge and screw dislocations is more pronounced.

## 1. Introduction

Accounting for the elastic anisotropy of the medium is essential in studying several physical phenomena related to the mechanics of crystal dislocations. Elastic anisotropy has strong effects on dislocation behavior in irradiated materials, the microscopic deformation of crystals, the diffusion characteristics of atomic species, and on coupled elasto-magneto-electric phenomena. For example, elastic calculations in neutron-irradiated iron and copper (Seif and Ghoniem, 2013) have shown that anisotropy plays a significant role in the calculation of *bias factors*, a measure of the preferential absorption by dislocations of self interstitial atoms rather than vacancies, which ultimately results in swelling and material deterioration (Greenwood et al., 1959; Wolfer, 2007). Similarly, elastic anisotropy is known to alter stress-driven diffusion pathways in crystals (Larcht'e and Cahn, 1982), and in particular solute diffusion near dislocations (Garnier et al., 2013). This process finds practical applications in defect *gettering* in semiconductors (Perevostchikov and Skoupov, 2005), and may be relevant to improving the efficiency of Si-based solar cells for the photovoltaic industry (Ziebarth et al., 2015). Anisotropy is also responsible for particular features observed in the high temperature deformation of $\alpha$-Fe, such as the presence of sharp corners in the shape of prismatic and glide loops (Fitzgerald and Yao, 2009; Fitzgerald and Aubry, 2010; Aubry et al., 2011), and the dependence of the critical stress necessary to bow out Frank-Read sources on its glide plane and orientation (Fitzgerald et al., 2012). The coupled elastic, electric and magnetic fields produced by an arbitrary three-dimensional dislocation loop in general anisotropic magneto-electro-elastic materials was developed by Han and Pan (2013). As a result of the coupling between the elastic, electric and magnetic fields, it was found that a dislocation, an electric potential discontinuity, or a magnetic potential discontinuity can induce simultaneous elastic, electric and magnetic fields. The coupling effect was also found to be very strong near the dislocation core. The elastic interaction between Self-Interstitial (SIA) Clusters, and between clusters and dislocations was included in kinetic Monte Carlo (KMC) simulations of damage evolution in irradiated bcc metals (Wen et al., 2005). The results indicate that near the dislocation core, SIA clusters, which normally migrate by 1D glide, rotate due to their elastic interactions. This rotation was found to be necessary to explain experimentally-observed dislocation decoration and raft formation in neutron irradiated pure iron. The elastic field of a general dislocation loop in anisotropic crystals was determined by incorporating numerically evaluated derivatives of Green's functions in the fast sum method (Han et al., 2003). The study showed that elastic anisotropy plays an important role in many dislocation mechanisms, such as local Peach-Koehler and self-forces, the operation of Frank-Read sources, dipole formation and break-up, and finally on dislocation junction strength. Compared to isotropic calculations, anisotropy was also found to yield image forces with opposite sign for dislocation models of bi-metal interface shearing (Chu et al., 2013), and therefore it may be important in predicting dislocation interactions with grain boundaries (e.g. Zbib et al., 2011; Wang et al., 2014). The elastic fields of three dimensional dislocation loops across anisotropic multilayer materials were studied by Han and Ghoniem (2005) and Ghoniem and Han (2005), while Zhang et al. (2016) considered the effects of core-spreading dislocation in anisotropic bi-materials.

Investigating these phenomena in the context of discrete dislocation dynamics (DDD) simulations requires the use of the anisotropic elastic theory of dislocations (e.g Mura, 1987). Compared to the isotropic case, the numerical implementation of the anisotropic elastic theory of dislocations suffers two main drawbacks, both of which are ultimately related to the Green's tensor of the anisotropic Navier differential operator (Lifshitz and Rosenzweig, 1947; Synge, 1957). The first drawback consists in the increased computational cost of the Green's tensor, which in classical anisotropic elasticity requires an additional integration on the equatorial circle of the unit sphere in Fourier space. In order to mitigate the heavy computational costs of Green's tensor, Aubry and Arsenlis (2013) have recently proposed an expansion in spherical harmonics, while Bertin et al. (2015) have proposed a fast Fourier Transform (FFT) method applicable to periodic dislocation systems. The second issue is due to the fact that the classical elastic fields of dislocations inherit the singularity of the Green's tensor at the origin, and therefore their numerical implementation requires some form of regularization. Although regularized theories of three-dimensional dislocation loops exist for isotropic materials (Cai et al., 2006; Banerjee et al., 2007; Lazar, 2012, 2013; Po et al., 2014), their anisotropic extension is non-trivial. In fact, their common regularization technique consists in removing the singularity from the derivatives of the distance function $R$ appearing in the Green's tensor by convolution with isotropic regularization functions. However, not only the anisotropic Green's tensor does not involve the derivatives of $R$ explicitly, but the very assumption that in anisotropic materials the regularization function remains isotropic is questionable.

In this paper we develop a non-singular theory of three-dimensional dislocation loops in anisotropic media. The theory is derived within a simplified version of Mindlin's anisotropic strain-gradient elasticity with up to six length scale parameters, a framework already introduced by Lazar and Po (2015b) and called Mindlin's anisotropic gradient elasticity with separable weak non-locality (see also Lazar and Po, 2015a; Seif et al., 2015). In recent years, the use of non-locality as a mean to describe the elastic fields of defects cores has received renewed attention (e.g. Taupin et al., 2014, 2017). In the proposed framework, the non-local parameters appear as the coefficients of an anisotropic Helmholtz differential operator. The Green's function of such operator is taken as the spreading function for the Burgers vector density. This choice is based on the ansatz that the volume affected by the plastic distortion is determined by the characteristic length scale parameters of the elastic continuum. This choice is justified by several considerations. First, the Green's function possesses unit integral over infinite space, and therefore it does not change the total Burgers vector density, a necessary condition for the non-singular fields to converge to their classical counterparts a few characteristic lengths away from the dislocation core. Second, because atomistic calculations show that such length scale parameters are in the order of nearest-neighbor interatomic distances, the plastically distorted volume remains localized between interatomic planes. Third, because the anisotropic Helmholtz operator is defined by a symmetric tensor of rank two, which must be invariant under material symmetry operations, the theory entails spreading functions which possess a different number of length scale parameters for different classes of material symmetry. In particular the spreading function possesses six independent length scales in triclinic, four in monoclinic, three in orthorhombic, two in tetragonal, hexagonal, and trigonal, and one for cubic crystals. For low symmetry crystals, therefore, the characteristic length scales of the "core regularization" are different along different crystallographic directions. Fourth, regardless of the class of material symmetry, the proposed non-singular dislocation theory regularizes the classical one maintaining formally identical dislocation key formulas. As a matter of fact, all the non-singular anisotropic dislocation fields can formally be obtained from their classical counterparts by replacing the singular Green's tensor with its regularized version given in Lazar and Po (2015b). The anisotropic non-singular versions of all classical dislocation equations are derived, and in particular: both Volterra and Burgers equations for displacement (Volterra, 1907; Burgers, 1939a,b), the Mura-Willis distortion equation (Mura, 1969; Willis, 1967), the Peach-Koehler stress and force equations (Peach and Koehler, 1950), and Blin's formula for the interaction energy between two dislocation loops (Blin, 1955). Apart from the generalized solid angle appearing in the displacement field, all equations are expressed in terms of non-singular line integrals, and therefore they are well-suited for applications in DDD simulations. An interesting technical remark is that the interaction energy equation is derived without the use of a stress function, a result which, to the best of our knowledge, was never obtained before.

The paper is organized as follows. In section 2 we develop the anisotropic non-singular theory of three dimensional dislocation loops, as a systematic generalization of the classical anisotropic theory. In section 3 we provide a physical interpretation of the simplified Mindlin framework, and we propose methods to determine the tensor of gradient length scale parameters for several crystal structures. In section 4 molecular statics calculations of the stress field in both cubic and hexagonal crystals are compared to the proposed anisotropic non-singular dislocation theory. Elastic calculations are performed using the DDD method, that is by numerical line integration of the proposed elastic kernels along the dislocation lines. Moreover, because the length-scale parameters are found by a deterministic method, the comparison is made without any fitting procedure. In section 5 we discuss the topic of self energy of dislocation loops, and we provide several results useful to compare non-singular anisotropic, non-singular isotropic, and classical theories. Finally, discussion and conclusions are presented in section 6.

## 2. The eigendistortion problem in anisotropic elasticity

The objective of this section is to derive all the fundamental equations of dislocation theory in Mindlin's anisotropic gradient elasticity with separable weak non-locality, and compare them with their counterparts in classical anisotropic elasticity (e.g. Mura, 1987). Both classical and gradient anisotropic theories of dislocation loops can be understood in the general framework of linearized incompatible elasticity. The main kinematic assumption of this framework is that the displacement gradient $\boldsymbol{u}$ is split additively into an elastic distortion $\boldsymbol{\beta}$ and a plastic distortion $\boldsymbol{\beta}^P$:

$$u_{i,j} = \beta_{ij} + \beta_{ij}^P. \tag{1}$$

In this kinematic framework, plastic distortion is caused by dislocations, and the particular form of the tensor $\boldsymbol{\beta}^P$ will be specified later for the classical theory and for the gradient theory. Given a specific form of $\boldsymbol{\beta}^P$, we recall that the corresponding dislocation density tensor $\boldsymbol{\alpha}$ is obtained as the negative curl of $\boldsymbol{\beta}^P$:

$$\alpha_{ij} = -\epsilon_{jkm}\beta_{im,k}^P. \tag{2}$$

In the following we shall derive the fundamental equations of dislocation loops in classical and gradient elasticity. Clearly all the classical results are well-known, but their derivation is summarized here to offer a guideline for the corresponding concepts in gradient elasticity, and to allow a one-to-one comparison of the fundamental equations. Moreover, because dislocation theory is a particular instance of eigendistortion theory, specialized for a given form of $\boldsymbol{\beta}^P$, results are derived for the general eigendistortion problem first, and then specialized for dislocation loops. Such an approach will be followed for both the classical case (sections 2.1 and 2.2) and the gradient case (sections 2.3 and 2.4).

Before discussing the eigendistortion problem in classical and gradient elasticity, we remark that, if the elastic distortion $\boldsymbol{\beta}$ induced by the plastic eigendistortion $\boldsymbol{\beta}^P$ is known, then the displacement field $\boldsymbol{u}$ can be expressed using the following strategy (Lazar and Kirchner, 2013), valid in both classical and gradient elasticity. First we take the derivative of Eq. (1) with respect to $x_j$ in order to obtain the Poisson equation:

$$\Delta u_i = u_{i,jj} = \beta_{ij,j} + \beta^P_{ij,j}, \tag{3}$$

where $\Delta = \partial_j \partial_j$ is the Laplace operator. Second we "invert" Eq. (3) using the Green's function[1] of the Laplace operator $G^\Delta$:

$$u_i = \left(\beta^P_{ij,j} + \beta_{ij,j}\right) * G^\Delta = \beta^P_{ij} * G^\Delta_{,j} + \beta_{ij,j} * G^\Delta. \tag{5}$$

In Eq. (5) we have introduced the symbol $*$ to indicate convolution over the infinite three-dimensional space. It will be clearer in the following sections that Eq. (5) can be considered as a *generalized Burgers equation*. In fact, although Eq. (5) is valid for an arbitrary source of plastic eigendistortion in an anisotropic medium, its name is well-justified in the case of dislocation loops because the term $\beta^P_{ij} * G^\Delta_{,j}$ corresponds to the contribution of the solid angle subtended by a loop, which was first isolated by Burgers (Burgers, 1939a,b) in the classical isotropic case.

## 2.1. The eigendistortion problem in classical anisotropic elasticity

The standard linear elastic medium can be characterized by a strain energy density, $W$, expressed as a quadratic form of the elastic strain[2] $\varepsilon^0_{ij}$:

$$W = \frac{1}{2} \mathbb{C}_{ijkl} \varepsilon^0_{ij} \varepsilon^0_{kl}, \tag{6}$$

where $\mathbb{C}_{ijkl}$ is the standard rank-4 tensor of elastic moduli. Given $W$, in the absence of body forces the equilibrium equation reads:

$$\left(\frac{\partial W}{\partial \varepsilon^0_{ij}}\right)_{,j} = \sigma^0_{ij,j} = 0, \tag{7}$$

where

$$\sigma^0_{ij} = \frac{\partial W}{\partial \varepsilon^0_{ij}} = \mathbb{C}_{ijkl} \varepsilon^0_{kl} = \mathbb{C}_{ijkl} \beta^0_{kl} \tag{8}$$

is the Cauchy stress tensor. Using the additive decomposition (1), the equilibrium equation in terms of displacement takes the form of the following inhomogeneous Navier equation:

$$L_{ik} u^0_k = \mathbb{C}_{ijkl} \beta^{P,0}_{kl,j}, \tag{9}$$

where

$$L_{ik} = \mathbb{C}_{ijkl} \partial_j \partial_l \tag{10}$$

is the Navier differential operator. This operator admits the well-known Green's tensor $G^0_{im}$ given in Eq. (A.4). Because the Green's tensor $G^0_{im}$ is the fundamental solution of the Navier operator, the particular solution of Eq. (9) can be obtained from it by convolution with the source term. In particular, for an infinite medium the displacement field reads (Mura, 1987):

$$u^0_i = -\mathbb{C}_{mnpq} G^0_{im,n} * \beta^{P,0}_{pq} \qquad \text{(generalized Volterra equation)}. \tag{11}$$

---

[1] The three-dimensional Green's function of the Laplace operator is (Vladimirov, 1971)

$$G^\Delta(\mathbf{R}) = -\frac{1}{4\pi R} \tag{4}$$

where $R = \sqrt{\mathbf{R} \cdot \mathbf{R}}$ is the Euclidean norm of $\mathbf{R}$.

[2] We use the superscript 0 to indicate classical fields.

With dislocation theory in mind, Eq. (11) can be considered a *generalized Volterra solution* valid for an arbitrary source of eigendistortion. However we are interested in obtaining an alternative expression of the displacement field in the form of Eq. (5), that is a *generalized Burgers solution*, where the contribution of elastic and plastic distortions appear in separate additive terms. To obtain such result, we first find the elastic distortion from Eqs. (1) and (2) using the Mura-Willis procedure[3]:

$$\beta_{ij}^0 = \mathbb{C}_{mnpq} \epsilon_{jqr} G_{im,n}^0 * \alpha_{pr}^0 \qquad \text{(Mura-Willis equation)} \qquad (13)$$

where $\alpha_{ij}^0 = -\epsilon_{jkm}\beta_{im,k}^{P,0}$. Then, substituting Eq. (13) in (5), we find the generalized Burgers solution for the displacement field (see also Lazar and Kirchner (2013)):

$$u_i^0 = G_{,j}^\Delta * \beta_{ij}^{P,0} - \mathbb{C}_{mnpq} \epsilon_{jqr} F_{jnim}^0 * \alpha_{pr}^0 \qquad \text{(generalized Burgers equation)}. \qquad (14)$$

Note that, following Kirchner (1984) and Lazar and Kirchner (2013), in Eq. (14) we have introduced the fourth-rank tensor $F_{jnim}^0$ defined as:

$$F_{jnim}^0 = -G_{im,jn}^0 * G^\Delta. \qquad (15)$$

Similar to the Green's tensor, also the "$F$-tensor" has an explicit form, which is given in Eq. (A.7). The $F$-tensor is also useful in deriving a compact expression for the interaction energy between two sources of eigendistortion. In fact, the elastic interaction energy between two sources of eigendistortion, labeled A and B respectively, can be obtained as[4]:

$$W_{AB} = \int_{\mathbb{R}^3} \epsilon_{jkl} \mathbb{C}_{ilmn} \epsilon_{npq} \mathbb{C}_{rstp} \left( F_{skmr}^0 * \alpha_{tq}^{0(A)} \right) \alpha_{ij}^{0(B)} \, \text{d}V \qquad \text{(generalized Blin's interaction energy equation)}. \qquad (17)$$

To conclude this summary, we recall that the Cauchy stress field $\sigma_{ij}^0$ can be obtained using the Hooke's law (8) and the elastic distortion (13)

$$\sigma_{ij}^0 = \mathbb{C}_{ijkl} \mathbb{C}_{mnpq} \epsilon_{lqr} G_{km,n}^0 * \alpha_{pr}^0 \qquad \text{(anisotropic Peach-Koehler stress equation)}, \qquad (18)$$

---

[3] Starting from Eq. (11), in our notation the Mura-Willis procedure reads:

$$\begin{aligned}
u_{i,j}^0 &= -\mathbb{C}_{mnpq} G_{im,n}^0 * \beta_{pq,j}^{P,0} = -\mathbb{C}_{mnpq} G_{im,n}^0 * \left( \beta_{pj,q}^{P,0} + \epsilon_{qjr} \alpha_{pr} \right) = -\mathbb{C}_{mnpq} G_{im,nq}^0 * \beta_{pj}^{P,0} - \mathbb{C}_{mnpq} \epsilon_{qjr} G_{im,n}^0 * \alpha_{pr} \\
&= -L_{pm} G_{im} * \beta_{pj}^{P,0} - \mathbb{C}_{mnpq} \epsilon_{qjr} G_{im,n}^0 * \alpha_{pr} = \delta_{ip} \delta * \beta_{pj}^{P,0} - \mathbb{C}_{mnpq} \epsilon_{qjr} G_{im,n}^0 * \alpha_{pr} = \beta_{ij}^{P,0} + \mathbb{C}_{mnpq} \epsilon_{jqr} G_{im,n}^0 * \alpha_{pr}
\end{aligned} \qquad (12)$$

The elastic distortion $\beta$ follows from (3).

[4] In order to prove (17), we use the following representation of the gradient of the Green's tensor in terms of the $F$-tensor:

$$G_{im,j}^0 = G_{im,jnn}^0 * G^\Delta = -F_{jnim,n}^0. \qquad (16)$$

With this, a derivation of (17) goes as follows:

$$\begin{aligned}
W_{AB} &= \int_{\mathbb{R}^3} \sigma_{il}^{0(A)} \beta_{il}^{0(B)} \, \text{d}V = \int_{\mathbb{R}^3} \mathbb{C}_{ilmn} \epsilon_{npq} \mathbb{C}_{rstp} \left( G_{mr,s}^0 * \alpha_{tq}^{0(A)} \right) \beta_{il}^{0(B)} \, \text{d}V = -\int_{\mathbb{R}^3} \mathbb{C}_{ilmn} \epsilon_{npq} \mathbb{C}_{rstp} \left( F_{skmr,k}^0 * \alpha_{tq}^{0(A)} \right) \beta_{il}^{0(B)} \, \text{d}V \\
&= \int_{\mathbb{R}^3} \mathbb{C}_{ilmn} \epsilon_{npq} \mathbb{C}_{rstp} \left( F_{skmr}^0 * \alpha_{tq}^{0(A)} \right) \beta_{il,k}^{0(B)} \, \text{d}V = \int_{\mathbb{R}^3} \mathbb{C}_{ilmn} \epsilon_{npq} \mathbb{C}_{rstp} \left( F_{skmr}^0 * \alpha_{tq}^{0(A)} \right) \left( \beta_{ik,l}^{0(B)} + \epsilon_{jkl} \alpha_{ij}^{0(B)} \right) \text{d}V \\
&= -\int_{\mathbb{R}^3} \underbrace{\mathbb{C}_{ilmn} \epsilon_{npq} \mathbb{C}_{rstp} \left( G_{mr,skl}^0 * \alpha_{tq}^{0(A)} \right)}_{\sigma_{il,lk}} * G^\Delta \beta_{ik}^{0(B)} \, \text{d}V + \int_{\mathbb{R}^3} \epsilon_{jkl} \mathbb{C}_{ilmn} \epsilon_{npq} \mathbb{C}_{rstp} \left( F_{skmr}^0 * \alpha_{tq}^{0(A)} \right) \alpha_{ij}^{0(B)} \, \text{d}V \\
&= \int_{\mathbb{R}^3} \epsilon_{jkl} \mathbb{C}_{ilmn} \epsilon_{npq} \mathbb{C}_{rstp} \left( F_{skmr}^0 * \alpha_{tq}^{0(A)} \right) \alpha_{ij}^{0(B)} \, \text{d}V.
\end{aligned}$$

This result shows that the interaction energy between two sources of eigendistortion depends only on their curls. Therefore it can be regarded as an anisotropic generalization of Blin's equation (Blin, 1955), valid for arbitrary sources of eigendistortion. Surprisingly, such fundamental result has long been missing in anisotropic eigendistortion theory, and it was found only recently by Lazar and Kirchner (2013) using the method of stress functions. To our knowledge, the derivation above is the first without stress functions.

while the configurational force exerted on the eigendistortion can be computed from the divergence of Eshelby's stress tensor. Illustrating this general procedure will turn out to be useful later for the gradient case, for which the definition of the force is not self-evident. For the classical theory, this clearly gives the Peach-Koehler force

$$\mathcal{F}_k^0 = \int_{\mathbb{R}^3} \left( W \delta_{kj} - \sigma_{ij}^0 \beta_{ik}^0 \right)_{,j} \mathrm{d}V = \int_{\mathbb{R}^3} \epsilon_{kjm} \sigma_{ij}^0 \alpha_{im}^0 \mathrm{d}V \,. \tag{19}$$

### 2.2. Dislocation loops in classical anisotropic elasticity

The key equations of classical dislocation theory are now obtained by specializing the form of the eigendistortion tensor $\boldsymbol{\beta}^P$. For a dislocation extending over a surface $\mathcal{S}$, the classical plastic eigendistortion tensor is taken in following form:

$$\beta_{kl}^{P,0}(\boldsymbol{x}) = -\int_{\mathcal{S}} \delta(\boldsymbol{x} - \boldsymbol{x}') b_k \, \mathrm{d}A_l' \,, \tag{20}$$

where $\delta$ is the Dirac delta function, and $b_i$ is the displacement jump across $\mathcal{S}$, or Burgers vector. For a Volterra dislocation, as opposed to a Somigliana dislocation, the Burgers vector is constant. In this case, the tensor $\alpha_{ij}^0 = -\epsilon_{jkm} \beta_{im,k}^{P,0}$ turns out to be concentrated on the dislocation line, that is the closed line $\mathcal{L} = \partial \mathcal{S}$ bounding the surface $\mathcal{S}$, and it is known as the *dislocation density tensor*:

$$\alpha_{ij}^0(\boldsymbol{x}) = \oint_{\mathcal{L}} \delta(\boldsymbol{x} - \boldsymbol{x}') b_i \, \mathrm{d}L_j' \,. \tag{21}$$

Using the special forms (20) and (21) in the results obtained in the previous section we find all the key equations of classical dislocation theory in the anisotropic case. Letting $\boldsymbol{R} = \boldsymbol{x} - \boldsymbol{x}'$, these are:

$$u_i^0(\boldsymbol{x}) = -\frac{b_i \Omega^0(\boldsymbol{x})}{4\pi} - \oint_{\mathcal{L}} \mathbb{C}_{mnpq} \epsilon_{jqr} b_p F_{jnim}^0(\boldsymbol{R}) \, \mathrm{d}L_r' \qquad \text{(anisotropic Burgers equation)} \tag{22a}$$

$$\beta_{ij}^0(\boldsymbol{x}) = \oint_{\mathcal{L}} \mathbb{C}_{mnpq} \epsilon_{jqr} G_{im,n}^0(\boldsymbol{R}) b_p \, \mathrm{d}L_r' \qquad \text{(Mura-Willis equation)} \tag{22b}$$

$$\sigma_{ij}^0(\boldsymbol{x}) = \oint_{\mathcal{L}} \mathbb{C}_{ijkl} \mathbb{C}_{mnpq} \epsilon_{lqr} G_{km,n}^0(\boldsymbol{R}) b_p \, \mathrm{d}L_r' \qquad \text{(anisotropic Peach-Koehler stress equation)} \tag{22c}$$

$$W_{AB} = \oint_{\mathcal{L}_A} \oint_{\mathcal{L}_B} \epsilon_{jkl} \mathbb{C}_{ilmn} \epsilon_{npq} \mathbb{C}_{rstp} F_{skmr}^0(\boldsymbol{R}) \, b_t^A b_i^B \, \mathrm{d}L_q^A \, \mathrm{d}L_j^B \qquad \text{(anisotropic Blin's formula)} \tag{22d}$$

$$\mathcal{F}_k^0 = \oint_{\mathcal{L}} \epsilon_{kjm} \sigma_{ij}^0 b_i \, \mathrm{d}L_m \qquad \text{(Peach-Koehler force)} \tag{22e}$$

Note that, in the Burgers equation, we have introduced the solid angle $\Omega^0(\boldsymbol{x})$ subtended by the loop from the relationship $G_{,j}^\Delta * \beta_{ij}^{P,0} = -b_i \Omega^0/4\pi$, which yields:

$$\Omega^0(\boldsymbol{x}) = 4\pi \int_{\mathcal{S}} G_{,j}^\Delta \, \mathrm{d}A_j' = \int_{\mathcal{S}} \frac{R_j}{R^3} \, \mathrm{d}A_j' \,. \tag{23}$$

### 2.3. The eigendistortion problem in anisotropic gradient elasticity

As a generalization of classical elasticity, we shall now consider the eigendistortion problem in incompatible Mindlin's form II anisotropic gradient elasticity (Mindlin, 1964, 1972). In such a framework, the strain energy density of a centrosymmetric[5] material reads:

$$W = \frac{1}{2} \mathbb{C}_{ijkl} \beta_{ij} \beta_{kl} + \frac{1}{2} \mathbb{D}_{ijmkln} \beta_{ij,m} \beta_{kl,n} \,, \tag{24}$$

---
[5] Due to the assumption of centrosymmetry, there is no coupling between $\varepsilon_{ij}$ and $\varepsilon_{ij,m}$.

where $\mathbb{D}_{ijmkln}$ is a sixth-rank tensor of strain gradient coefficients with symmetries $\mathbb{D}_{ijmkln} = \mathbb{D}_{jimkln} = \mathbb{D}_{ijmlkn} = \mathbb{D}_{klnijm}$. Following Lazar and Po (2015a,b), we assume that the following assumption holds

$$\mathbb{D}_{ijmkln} \approx \mathbb{C}_{ijkl}\Lambda_{mn}, \tag{25}$$

that is, $\mathbb{D}_{ijmkln}$ is multiplicatively split into the fourth-rank tensor of material moduli, $\mathbb{C}_{ijkl}$, and a second-rank tensor of gradient length scale parameters with units of squared length, $\Lambda_{mn}$. From a physical viewpoint, the decomposition (25) represents the separation of the two sources of anisotropy present in Mindlin's anisotropic gradient elasticity, namely the elastic bulk anisotropy and the anisotropy of the gradient length scale parameters, which models a weak nonlocal anisotropy[6]. The determination of the tensor $\mathbf{\Lambda}$ and the physical interpretation of the decomposition (25) are further discussed in section 3.

Given $W$ and the aforementioned assumption of linearized kinematics, the principle of virtual work demands that the displacement field, $u_k$, satisfies the following equilibrium equation:

$$\left(\frac{\partial W}{\partial \varepsilon_{ij}} - \left(\frac{\partial W}{\partial \varepsilon_{ij,m}}\right)_{,m}\right)_{,j} = \left(\sigma_{ij} - \tau_{ijm,m}\right)_{,j} = 0, \tag{26}$$

where

$$\sigma_{ij} = \frac{\partial W}{\partial \varepsilon_{ij}} = \mathbb{C}_{ijkl}\beta_{kl} \tag{27}$$

$$\tau_{ijm} = \frac{\partial W}{\partial \varepsilon_{ij,m}} = \mathbb{C}_{ijkl}\Lambda_{mn}\beta_{kl,n} = \Lambda_{mn}\sigma_{ij,n} \tag{28}$$

are the Cauchy stress tensor and the double stress tensor, respectively. By virtue of the additive decomposition (1), the equilibrium equation in terms of displacement takes the form of the following inhomogeneous Navier-Helmholtz equation

$$L_{ik}Lu_k = \mathbb{C}_{ijkl}L\beta^P_{kl,j}, \tag{29}$$

where

$$L = 1 - \Lambda_{mn}\partial_m\partial_n, \tag{30}$$

is an anisotropic Helmholtz operator.

The source of eigendistortion $\boldsymbol{\beta}^P$ is assumed to be a "regularized" version of the classical one, obtained by convolution with an appropriate "spreading function". As already discussed in the introduction, the Green's function of the anisotropic Helmholtz operator[7] plays here the role of the "spreading function". Therefore, the regularized eigendistortion is:

$$\beta^P_{ij} = \beta^{P,0}_{ij} * G^L \qquad \text{or} \qquad L\beta^P_{ij} = \beta^{P,0}_{ij}. \tag{32}$$

As a consequence of (32) and (2), we obtain

$$\alpha_{ij} = \alpha^0_{ij} * G^L \qquad \text{or} \qquad L\alpha_{ij} = \alpha^0_{ij}. \tag{33}$$

---

[6] A similar decomposition into bulk anisotropy and nonlocal anisotropy is also present in nonlocal anisotropic elasticity (see, e.g., Lazar and Agiasofitou (2011)).

[7] The Green's function $G^L$ of the anisotropic Helmholtz operator (30) satisfies $LG^L = \delta$, and it has the following representation (c.f. Lazar and Po, 2015b):

$$G^L(\boldsymbol{R}) = \frac{1}{4\pi\sqrt{\det(\boldsymbol{\Lambda})}} \frac{e^{-\sqrt{\boldsymbol{R}^T\boldsymbol{\Lambda}^{-1}\boldsymbol{R}}}}{\sqrt{\boldsymbol{R}^T\boldsymbol{\Lambda}^{-1}\boldsymbol{R}}}. \tag{31}$$

Note that, in the case of dislocations, the spreading of $\boldsymbol{\alpha}$ corresponds to the well-known idea of distributing the singular Burgers vector density in the volume around the dislocation line (e.g. de Wit, 1960; Lothe, 1992; Cai et al., 2006; Lazar, 2012, 2013; Po et al., 2014). However, "spreading functions" used in the past possess an isotropic character. Here, on the other hand, the "spreading function" is anisotropic, because the tensor $\boldsymbol{\Lambda}$ possesses a different number of independent length scale parameters depending on the symmetry class of the crystal (Lazar and Po, 2015b). This concept is further discussed in 2.4.

With Eq. (32), the equilibrium equation becomes

$$L_{ik} L u_k = \mathbb{C}_{ijkl} \beta^{P,0}_{kl,j}. \tag{34}$$

Its solution reads

$$u_i = -\mathbb{C}_{mnpq} G_{im,n} * \beta^{P,0}_{pq} \qquad \text{(generalized Volterra equation)}, \tag{35}$$

where

$$G_{im} = G^0_{im} * G^L \tag{36}$$

is the non-singular Green's tensor of the twofold anisotropic Navier-Helmholtz operator $L_{ik} L$. The Green's tensor $G_{im}$ was recently determined by Lazar and Po (2015b), and it is given explicitly in Appendix B.

Note that Eq. (35) is the generalized Volterra solution for the eigendistortion problem in gradient elasticity. Similar to what done for the classical theory, we now seek the generalized Burgers solution. In order to obtain such solution we first need the elastic distortion tensor, which we find following the Mura-Willis procedure:

$$\beta_{ij} = \mathbb{C}_{mnpq} \epsilon_{jqr} G_{im,n} * \alpha^0_{pr} \qquad \text{(Mura-Willis equation)}. \tag{37}$$

Similar to the classical case, we now use Eq. (37) in (5), and obtain

$$u_i = \beta^P_{ij,j} * G^\Delta + \mathbb{C}_{mnpq} \epsilon_{jqr} G_{im,jn} * \alpha^0_{pr} * G^\Delta. \tag{38}$$

In order to simplify (38), we note the identity $\beta^P_{ij,j} * G^\Delta = \left(\beta^{P,0}_{ij} * G^L\right)_{,j} * G^\Delta = \left(G^\Delta * G^L\right)_{,j} * \beta^{P,0}_{ij} = G^{\Delta L}_{,j} * \beta^{P,0}_{ij}$, where $G^{\Delta L} = G^\Delta * G^L$ is the Green's tensor of the anisotropic Laplace-Helmholtz operator which suggests to introduce the Green's tensor of the one-fold anisotropic Laplace-Helmholtz operator, which is derived in Appendix C. In particular using $G^{\Delta L}_{,j}$ from Eq. (C.4), we obtain the generalized Burgers equation for the displacement field:

$$u_i = G^{\Delta L}_{,j} * \beta^{P,0}_{ij} - \mathbb{C}_{mnpq} \epsilon_{jqr} F_{jnim} * \alpha^0_{pr} \qquad \text{(generalized Burgers equation)}. \tag{39}$$

Note that, in analogy to the classical case, in (39) we have defined the $\boldsymbol{F}$-tensor in gradient elasticity as

$$F_{jnim} = -G_{im,jn} * G^\Delta. \tag{40}$$

The explicit representation of this tensor is given in Appendix B. Continuing the analogy with the classical case, the $\boldsymbol{F}$-tensor can be also used to find the interaction energy between two sources of eigendistortion. In fact, neglecting surface integrals that vanish at infinity, we have[8]:

$$W_{AB} = \int_{\mathbb{R}^3} \epsilon_{jkl} \mathbb{C}_{ilmn} \epsilon_{npq} \mathbb{C}_{rstp} \left( F_{skmr} * \alpha_{tq}^{0(A)} \right) \alpha_{ij}^{0(B)} \, \mathrm{d}V. \tag{41}$$

This result shows that, also in Mindlin's gradient elasticity with separable weak non-locality, the interaction energy between two sources of eigendistortion depends only on their curls.

Finally, the Cauchy stress field generated by the eigendistortion is found by using Eqs. (27) and (37)

$$\sigma_{ij} = \mathbb{C}_{ijkl} \mathbb{C}_{mnpq} \epsilon_{lqr} G_{km,n} * \alpha_{pr}^0 \qquad \text{(anisotropic Peach-Koehler stress equation)}, \tag{42}$$

while the configurational force exerted on the source of eigendistortion, is found from the divergence of Eshelby's stress tensor in gradient elasticity (Lazar and Kirchner, 2007). This yields the Peach-Koehler force valid in gradient elasticity:

$$\mathcal{F}_k = \int_{\mathbb{R}^3} \left[ W \delta_{kj} - \left( \sigma_{ij} - \tau_{ijm,m} \right) \beta_{ik} - \sigma_{im,j} \beta_{im,k} \right]_{,j} \mathrm{d}V = \int_{\mathbb{R}^3} \epsilon_{kjl} \left[ \sigma_{ji} \alpha_{il} + \tau_{ijm} \alpha_{il,m} \right] \mathrm{d}V = \int_{\mathbb{R}^3} \epsilon_{kjl} L(\sigma_{ji}) \alpha_{il} \, \mathrm{d}V$$

$$= \int_{\mathbb{R}^3} \epsilon_{kjl} \sigma_{ji} L \alpha_{il} \, \mathrm{d}V = \int_{\mathbb{R}^3} \epsilon_{kjl} \sigma_{ji} \alpha_{il}^0 \, \mathrm{d}V. \tag{43}$$

Note that in the last equality we have used Eq. (33), and integration by parts was employed to drop terms at infinity. Therefore, it is shown that, in Mindlin's gradient elasticity with separable weak non-locality, the configurational force force exerted on a source of eigendistortion retains its classical form, in the sense that it involves only the Cauchy stress.

### 2.4. Dislocation loops in anisotropic gradient elasticity

All the fundamental equations of dislocation theory are now obtained as a special case, by substituting the special forms (20) and (21) in the main results obtained in section 2.3. Letting $\boldsymbol{R} = \boldsymbol{x} - \boldsymbol{x}'$, and using the sifting property of the Dirac-$\delta$ function we obtain all the fundamental equations of dislocation theory in Mindlin's gradient elasticity with weak non-locality. These are:

$$u_i(\boldsymbol{x}) = -\frac{b_i \Omega(\boldsymbol{x})}{4\pi} - \oint_{\mathcal{L}} \mathbb{C}_{mnpq} \epsilon_{jqr} b_p F_{jnim}(\boldsymbol{R}) \, \mathrm{d}L'_r \qquad \text{(anisotropic Burgers equation)} \tag{44a}$$

$$\beta_{ij}(\boldsymbol{x}) = \oint_{\mathcal{L}} \mathbb{C}_{mnpq} \epsilon_{jqr} G_{im,n}(\boldsymbol{R}) b_p \, \mathrm{d}L'_r \qquad \text{(Mura-Willis equation)} \tag{44b}$$

$$\sigma_{ij}(\boldsymbol{x}) = \oint_{\mathcal{L}} \mathbb{C}_{ijkl} \mathbb{C}_{mnpq} \epsilon_{lqr} G_{km,n}(\boldsymbol{R}) b_p \, \mathrm{d}L'_r \qquad \text{(anisotropic Peach-Koehler stress equation)} \tag{44c}$$

$$W_{AB} = \oint_{\mathcal{L}_A} \oint_{\mathcal{L}_B} \epsilon_{jkl} \mathbb{C}_{ilmn} \epsilon_{npq} \mathbb{C}_{rstp} F_{skmr}(\boldsymbol{R}) \, b_t^A b_i^B \, \mathrm{d}L_q^A \, \mathrm{d}L_j^B \qquad \text{(anisotropic Blin's formula)} \tag{44d}$$

$$\mathcal{F}_k = \oint_{\mathcal{L}} \epsilon_{kjm} \sigma_{ij} b_i \, \mathrm{d}L_m \qquad \text{(Peach-Koehler force)} \tag{44e}$$

---

[8] The derivation of (41) without a stress function is as follows:

$$W_{AB} = \int_{\mathbb{R}^3} \left( \sigma_{il}^{(A)} \beta_{il}^{(B)} + \tau_{ilm}^{(A)} \beta_{il,m}^{(B)} \right) \mathrm{d}V = \int_{\mathbb{R}^3} \left( \sigma_{il}^{(A)} - \tau_{ilm,m}^{(A)} \right) \beta_{il}^{(B)} \, \mathrm{d}V = \int_{\mathbb{R}^3} L\left( \sigma_{il}^{(A)} \right) \beta_{il}^{(B)} \, \mathrm{d}V$$

$$= \int_{\mathbb{R}^3} L\left( \mathbb{C}_{ilmn} \epsilon_{npq} \mathbb{C}_{rstp} G_{mr,s} * \alpha_{tq}^{0(A)} \right) \beta_{il}^{(B)} \, \mathrm{d}V = -\int_{\mathbb{R}^3} L\left( \mathbb{C}_{ilmn} \epsilon_{npq} \mathbb{C}_{rstp} F_{skmr,k} * \alpha_{tq}^{0(A)} \right) \beta_{il}^{(B)} \, \mathrm{d}V$$

$$= \int_{\mathbb{R}^3} L\left( \mathbb{C}_{ilmn} \epsilon_{npq} \mathbb{C}_{rstp} F_{skmr} * \alpha_{tq}^{0(A)} \right) \beta_{il,k}^{(B)} \, \mathrm{d}V = \int_{\mathbb{R}^3} L\left( \mathbb{C}_{ilmn} \epsilon_{npq} \mathbb{C}_{rstp} F_{skmr} * \alpha_{tq}^{0(A)} \right) \left( \beta_{ik,l}^{(B)} + \epsilon_{jkl} \alpha_{ij}^{(B)} \right) \mathrm{d}V$$

$$= -\int_{\mathbb{R}^3} \underbrace{\left( L \mathbb{C}_{ilmn} \epsilon_{npq} \mathbb{C}_{rstp} G_{mr,skl}^0 * \alpha_{tq}^{0(A)} \right)}_{L\sigma_{il,lk} = 0} * G^{\Delta} \beta_{ik}^{(B)} \, \mathrm{d}V + \int_{\mathbb{R}^3} L\left( \epsilon_{jkl} \mathbb{C}_{ilmn} \epsilon_{npq} \mathbb{C}_{rstp} F_{skmr} * \alpha_{tq}^{0(A)} \right) \alpha_{ij}^{(B)} \, \mathrm{d}V$$

$$= \int_{\mathbb{R}^3} \epsilon_{jkl} \mathbb{C}_{ilmn} \epsilon_{npq} \mathbb{C}_{rstp} \left( F_{skmr} * \alpha_{tq}^{0(A)} \right) L \alpha_{ij}^{(B)} \, \mathrm{d}V = \int_{\mathbb{R}^3} \epsilon_{jkl} \mathbb{C}_{ilmn} \epsilon_{npq} \mathbb{C}_{rstp} \left( F_{skmr} * \alpha_{tq}^{0(A)} \right) \alpha_{ij}^{0(B)} \, \mathrm{d}V.$$

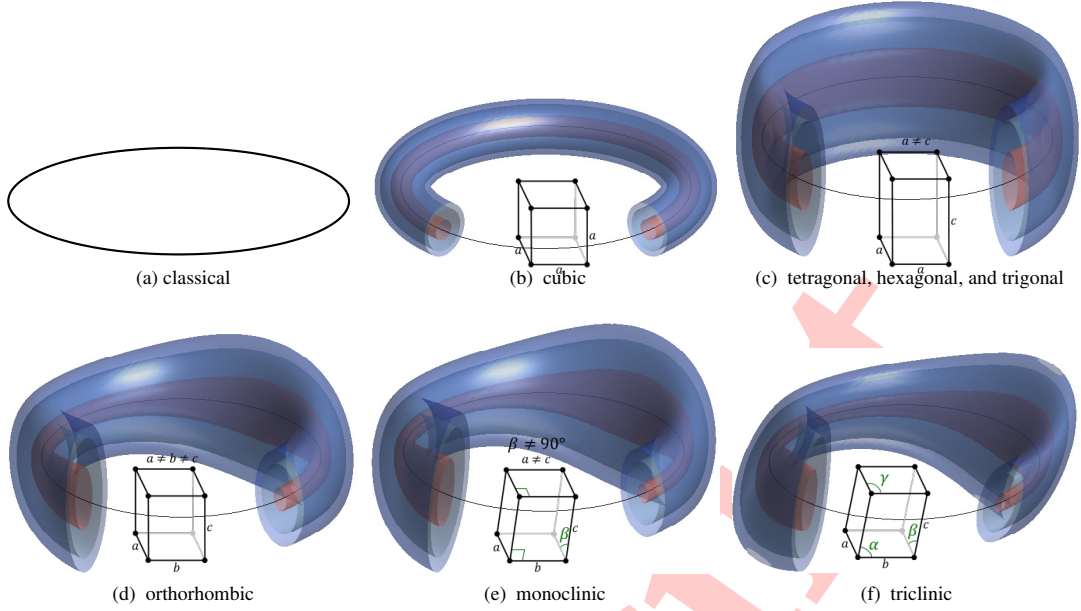

Figure 1: Burgers vector density for a circular dislocation loop, represented as isovalues of the norm of the dislocation density tensor $\boldsymbol{\alpha}$. (a) Classical elasticity, with tensor $\boldsymbol{\alpha} = \boldsymbol{\alpha}^0$ from (21). (b) Gradient elasticity, with tensor $\boldsymbol{\alpha}$ from (46), and cubic $\boldsymbol{\Lambda}$ (c) Gradient elasticity, with tensor $\boldsymbol{\alpha}$ from (46), and tetragonal (or hexagonal, or trigonal) $\boldsymbol{\Lambda}$. (d) Gradient elasticity, with tensor $\boldsymbol{\alpha}$ from (46), and orthorhombic $\boldsymbol{\Lambda}$. (e) Gradient elasticity, with tensor $\boldsymbol{\alpha}$ from (46), and monoclinic $\boldsymbol{\Lambda}$ (orientation 2‖$c$). (f) Gradient elasticity, with tensor $\boldsymbol{\alpha}$ from (46), and triclinic $\boldsymbol{\Lambda}$.

Note that, in the Burgers equation, we have introduced the generalized solid angle $\Omega(\boldsymbol{x})$ subtended by the loop from the relationship $G^{\Delta L}_{,j} * \beta^{P,0}_{ij} = -b_i\, \Omega/4\pi$, which yields

$$\Omega(\boldsymbol{x}) = 4\pi \int_{\mathcal{S}} G^{\Delta L}_{,j}(\boldsymbol{R})\, \mathrm{d}A'_j\,. \tag{44f}$$

As better discussed in Appendix D, the generalized solid angle admits a line integral representation which is useful for numerical implementation.

Note that, although never explicitly used in the elastic fields above, the distortion and dislocation density tensors of the loop are, respectively:

$$\beta^P_{kl}(\boldsymbol{x}) = -\int_{\mathcal{S}} G^L(\boldsymbol{x} - \boldsymbol{x}') b_k\, \mathrm{d}A'_l\,, \tag{45}$$

$$\alpha_{ij}(\boldsymbol{x}) = \oint_{\mathcal{L}} G^L(\boldsymbol{x} - \boldsymbol{x}') b_i\, \mathrm{d}L'_j\,. \tag{46}$$

A graphical representation of the Burgers vector density corresponding to (46) is shown in Fig. 1 for different crystal classes. For each class, the Burgers vector density distribution is controlled by the number of independent length scales parameters in the tensor $\boldsymbol{\Lambda}$, namely six for triclinic, four for monoclinic, three for orthorhombic, two for tetragonal, hexagonal, and trigonal, and one for cubic crystals (Lazar and Po, 2015b). Depending on the relative values of the components of $\boldsymbol{\Lambda}$, the Burgers vector density extends differently along different directions, and in general conforms to the lattice of the corresponding crystal class, with diagonal components of $\boldsymbol{\Lambda}$ determining its elongation along the three reference axes, and the off-diagonal components determining its tilting.

Before turning our attention to the numerical aspects of this work, we remark that, inspired by the theory developed in this section, it is possible to construct an approximate non-singular theory of dislocation loops in the classical framework, as better discussed in Appendix E.

## 3. Atomistic determination of the tensor of gradient length scale parameters $\mathbf{\Lambda}$

In order to complete the theory developed in the previous section, we need to formulate a criterion from which the tensor of gradient length scale parameters $\mathbf{\Lambda}$ can be determined. In this section we discuss the physical meaning of the decomposition (25), and we propose a method to estimate the tensor $\mathbf{\Lambda}$.

The full rank-six tensor $\mathbb{D}$ of strain-gradient coefficients can be obtained from atomistic calculations using the method proposed by Admal et al. (2017). The method yields explicit analytical expressions of all the elastic tensors in terms of the first and second derivatives of interatomic potentials with respect to relative distances between atoms, and referential relative vectors between them. The method can be considered the strain-gradient generalization of the atomistic representation of the classical tensor of elastic moduli $\mathbb{C}$ (Born and Huang, 1954; Tadmor and Miller, 2011). Therefore, given a certain interatomic potential, both tensors $\mathbb{C}$ and $\mathbb{D}$ can be computed self-consistently. With this premise, we propose two methods to compute the tensor of strain-gradient parameters $\mathbf{\Lambda}$.

### 3.1. The minimal residual method

The tensor $\mathbf{\Lambda}$ can be chosen such that it minimizes the $l^2$-norm of the difference $\mathbb{D} - \mathbb{C} \otimes \mathbf{\Lambda}$, that is:

$$\mathbf{\Lambda} = \arg\min \ \{\|\mathbb{D} - \mathbb{C} \otimes \mathbf{\Lambda}\|\} \,, \tag{47}$$

where it is understood that both tensors $\mathbb{D}$ and $\mathbb{C}$ are computed using their aforementioned atomistic representations for a certain material and interatomic potential. Note that the minimization (47) is subject to material symmetry constraints which limit the number of independent coefficients of the tensor $\mathbf{\Lambda}$ depending on the class of material symmetry (Lazar and Po, 2015b). For example, for a cubic crystal with symmetry planes parallel to the coordinate axes, the tensor $\mathbf{\Lambda}$ is diagonal, and the minimization argument yields

$$\Lambda_{ij}^{\text{cubic}} = \ell^2 \delta_{ij} \qquad \text{and} \qquad \ell^2 = \frac{1}{3} \frac{\mathbb{D}_{ijmklm}\mathbb{C}_{ijkl}}{\|C\|^2} \,. \tag{48}$$

For an orthorhombic crystal with symmetry planes parallel to the coordinate axes, we have

$$\Lambda_{ij}^{\text{orthorhombic}} = \text{diag}(\Lambda_{11}, \Lambda_{22}, \Lambda_{33}) \qquad \text{and} \qquad \Lambda_{\alpha\alpha} = \frac{\mathbb{D}_{ij\alpha kl\alpha}\mathbb{C}_{ijkl}}{\|C\|^2} \ (\text{no summation over } \alpha). \tag{49}$$

Similar relations can be obtained for all classes of materials symmetry. In the most general case (triclinic lattice) $\mathbf{\Lambda}$ is only required to be symmetric, and the minimization argument yields

$$\Lambda_{mn}^{\text{triclinic}} = \frac{1}{2} \frac{\left(\mathbb{D}_{ijmkln} + \mathbb{D}_{ijnklm}\right) \mathbb{C}_{ijkl}}{\|C\|^2} \,. \tag{50}$$

### 3.2. The projection method

The tensor $\mathbf{\Lambda}$ can also be estimated using the tensor $\mathbb{S} = \mathbb{C}^{-1}$, that is the inverse of the tensor of elastic moduli. This tensor is defined by the property (Teodosiu, 1982)

$$\mathbb{C}_{ijmn}\mathbb{S}_{mnkl} = \frac{1}{2}(\delta_{ik}\delta_{jl} + \delta_{il}\delta_{jk}) \,, \tag{51}$$

and therefore

$$\mathbb{C}_{ijkl}\mathbb{S}_{ijkl} = \mathbb{S}_{ijkl}\mathbb{C}_{ijkl} = 6 \,. \tag{52}$$

Multiplying both sides of (25) by $\mathbb{S}$, we obtain $\mathbb{S}_{ijkl}\mathbb{D}_{ijmkln} = 6\Lambda_{mn}$, and therefore, the tensor $\mathbf{\Lambda}$ can be given in terms of the two constitutive tensor $\mathbb{D}_{ijmkln}$ and $\mathbb{S}_{ijkl}$ as

$$\Lambda_{mn} = \frac{1}{6} \mathbb{S}_{ijkl}\mathbb{D}_{ijmkln} \,. \tag{53}$$

For instance, the tensor (53) reads for cubic crystals

$$\Lambda_{mn}^{\text{cubic}} = \ell^2 \delta_{mn} \qquad \text{and} \qquad \ell^2 = \frac{1}{18} \mathbb{S}_{ijkl}^{\text{cubic}} \mathbb{D}_{ijmklm}^{\text{cubic}}, \tag{54}$$

for hexagonal crystals

$$\begin{aligned}\Lambda_{mn}^{\text{hex}} &= \text{diag}(\Lambda_{11}, \Lambda_{11}, \Lambda_{33}) \\ &= \frac{1}{6} \text{diag}(\mathbb{S}_{ijkl}^{\text{hex}} \mathbb{D}_{ij1kl1}^{\text{hex}}, \mathbb{S}_{ijkl}^{\text{hex}} \mathbb{D}_{ij1kl1}^{\text{hex}}, \mathbb{S}_{ijkl}^{\text{hex}} \mathbb{D}_{ij3kl3}^{\text{hex}})\end{aligned} \tag{55}$$

and for orthorhombic crystals

$$\begin{aligned}\Lambda_{mn}^{\text{orhom}} &= \text{diag}(\Lambda_{11}, \Lambda_{22}, \Lambda_{33}) \\ &= \frac{1}{6} \text{diag}(\mathbb{S}_{ijkl}^{\text{orhom}} \mathbb{D}_{ij1kl1}^{\text{orhom}}, \mathbb{S}_{ijkl}^{\text{orhom}} \mathbb{D}_{ij2kl2}^{\text{orhom}}, \mathbb{S}_{ijkl}^{\text{orhom}} \mathbb{D}_{ij3kl3}^{\text{orhom}}).\end{aligned} \tag{56}$$

### 3.3. Physical interpretation of the assumption $D \approx C \otimes \Lambda$

The decomposition (25), that is $\mathbb{D} \approx \mathbb{C} \otimes \Lambda$, cannot be satisfied exactly, since there are many more independent material parameters in the tensor $\mathbb{D}$ (up to 171) compared to the tensor $\mathbb{C} \otimes \Lambda$ (up to 21+6). Therefore, the assumption (25) retains the status of an approximate constitutive material law. Nonetheless, the validity of such an approximation can be verified numerically. In order to do so, let us consider the Voigt representation of the tensors $\mathbb{D}$ (see Admal et al., 2017). In Voigt notation, the tensor $\mathbb{D}$ is represented as an $18 \times 18$ symmetric matrix $\mathcal{D}$. Its component $\mathcal{D}_{\alpha\beta}$ is the strain-energy contribution associated to strain-gradient components $\alpha$ and $\beta$. Here $\alpha$ and $\beta$ are Voigt indices, each being mapped to a triplet of tensor indices $\{ijk\}$ of the strain gradient tensor $\varepsilon_{ij,k}$, as follows:

$$\begin{aligned}&1 \mapsto 111, \quad 2 \mapsto 221, \quad 3 \mapsto 122, \quad 4 \mapsto 331, \quad 5 \mapsto 133, \quad 6 \mapsto 222, \\ &7 \mapsto 112, \quad 8 \mapsto 121, \quad 9 \mapsto 332, \quad 10 \mapsto 233, \quad 11 \mapsto 333, \quad 12 \mapsto 113, \\ &13 \mapsto 131, \quad 14 \mapsto 223, \quad 15 \mapsto 232, \quad 16 \mapsto 123, \quad 17 \mapsto 132, \quad 18 \mapsto 231.\end{aligned} \tag{57}$$

Figure 2a shows the matrix $\mathcal{D}$ for Al, using the interatomic potential of Ercolessi and Adams (1994). It can be observed that the matrix is diagonal dominated, with a few off-diagonal elements of significant importance compared to the diagonal ones. Similarly, Fig. 2b shows the Voigt representation of the tensor $\mathbb{C} \otimes \Lambda$ for the same material and interatomic potential, where $\Lambda$ is computed using the minimal residual method discussed above. By visual inspection, it is evident that the diagonal structure of the original tensor is maintained, although several off-diagonal components are zeroed out. From a physical viewpoint, therefore, the approximation $\mathbb{D} \approx \mathbb{C} \otimes \Lambda$ retains most of the energetic contribution of strain-gradient deformation modes acting individually (corresponding to the diagonal components of $\mathcal{D}$), but it tends to neglect the coupling energy between different deformation modes. Examples of strain-gradient deformation modes are illustrated in Fig. 2c-2f.

## 4. Comparison with Molecular Statics calculations

In this section we compare the anisotropic non-singular theory to Molecular Statics (MS) calculations of stress in both cubic and hexagonal crystals. For a given material and interatomic potential, the tensors $\mathbb{C}$ and $\mathbb{D}$ are first computed using their atomistic representations given in Admal et al. (2017). Using the same interatomic potential, an atomistic system is first relaxed at 0K. An infinite edge dislocation with line direction along the $x_3$ axis and Burgers vector along the $x_1$ axis is then introduced in the atomistic system by displacing atoms using the classical isotropic displacement solution

$$u_1(x_1, x_2) = \frac{b}{2\pi} \left[ \tan^{-1}\left(\frac{x_2}{x_1}\right) + \frac{1}{2(1-\nu)} \frac{x_1 x_2}{x_1^2 + x_2^2} \right], \tag{58a}$$

$$u_2(x_1, x_2) = \frac{b}{2\pi} \left[ \frac{1-2\nu}{4(1-\nu)} \ln\left(\frac{1}{x_1^2 + x_2^2}\right) + \frac{1}{2(1-\nu)} \frac{x_2^2}{x_1^2 + x_2^2} \right]. \tag{58b}$$

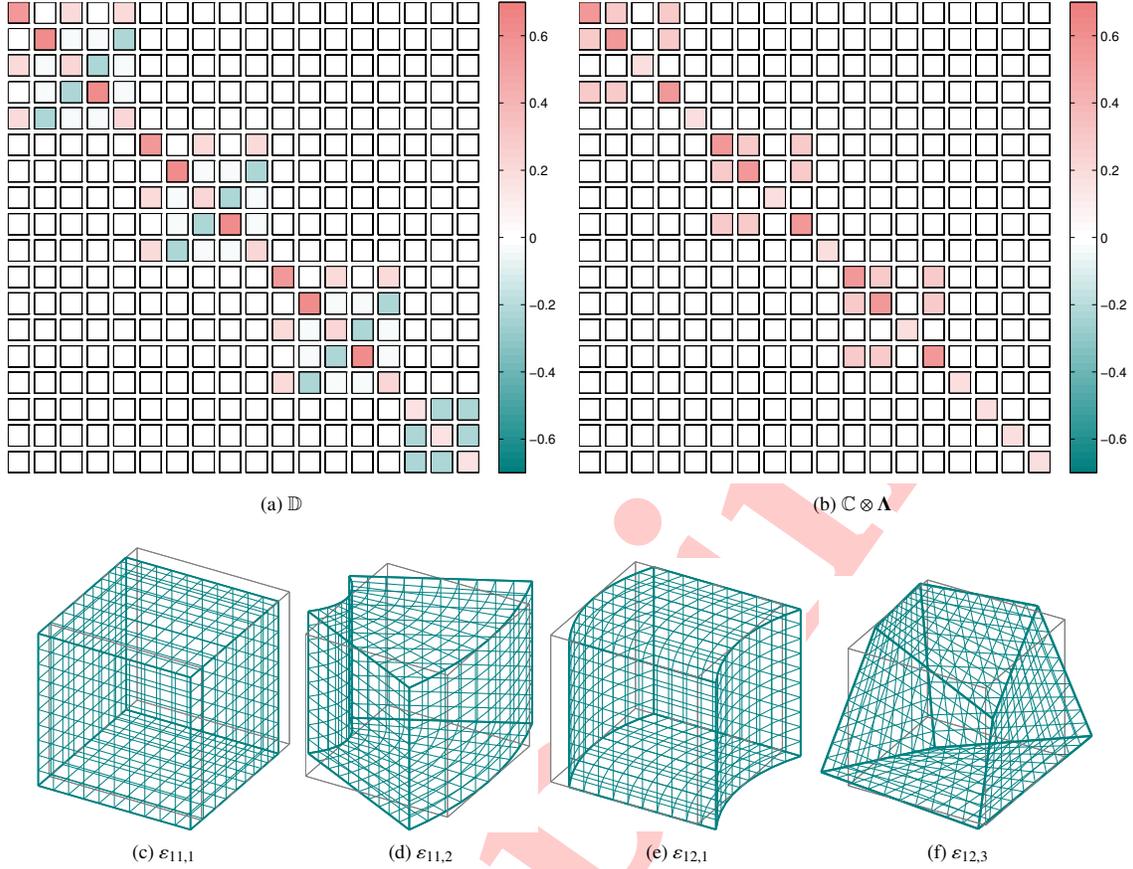

Figure 2: Graphical representation of the approximation $\mathbb{D} \approx \mathbb{C} \otimes \mathbf{\Lambda}$. (a) Voigt representation of the tensor $\mathbb{D}$ as a symmetric $18 \times 18$ matrix, computed using the method of Admal et al. (2017) for Al using the interatomic potential of Ercolessi and Adams (1994), in units of eV/Å. The block structure representation of the matrix is a general property of all cubic materials (Auffray et al., 2013). (b) Voigt representation of the tensor $\mathbb{C} \otimes \mathbf{\Lambda}$, where $\Lambda$ is computed with the minimal residual method discussed in section 3.1. (c)-(f) Examples of strain-gradient deformation modes of a continuum giving rise to strain-gradient contributions to the strain energy density.

The isotropic displacement field is used as a tentative configuration which is then further relaxed by conjugate gradient minimization of the total energy of the atomistic system. In the final relaxed configuration, the Cauchy stress is computed for each atom $\pi$ in the system from the expression

$$\boldsymbol{\sigma}^\pi = \frac{1}{2} \sum_{\substack{\alpha,\beta \\ \alpha \neq \beta}} \frac{\partial \mathcal{V}^\pi}{\partial r^{\alpha\beta}} \frac{\boldsymbol{r}^{\alpha\beta} \otimes \boldsymbol{r}^{\alpha\beta}}{r^{\alpha\beta}}, \tag{59}$$

where $\mathcal{V}^\pi$ is the potential energy of atom $\pi$, and $\boldsymbol{r}^{\alpha\beta} := \boldsymbol{x}^\alpha - \boldsymbol{x}^\beta$ is the vector joining the positions of atoms $\alpha$ and $\beta$ in the relaxed configuration, while $r^{\alpha\beta}$ is its norm. This quantity is then compared to the singular Cauchy stress (22c), and to the non-singular Cauchy stress (44c). The tensor of length scale parameters $\mathbf{\Lambda}$ appearing in the non-singular Cauchy stress is determined from tensors $\mathbb{C}$ and $\mathbb{D}$, as explained in section 3. Note that, because the length scale parameters are found in a deterministic way, no fitting procedure is used in the comparison between theoretical and atomistic results.

In order to demonstrate the feasibility of the non-singular equations in DDD applications, all elastic calculations were performed using the DDD method, that is by numerical integration of the stress kernel along the dislocation lines. To reduce the numerical cost of the Green tensor, the classical Green tensor is used beyond a distance $R \approx 5\|\mathbf{\Lambda}\|$ from the core region, without noticeable jumps in the stress profiles.

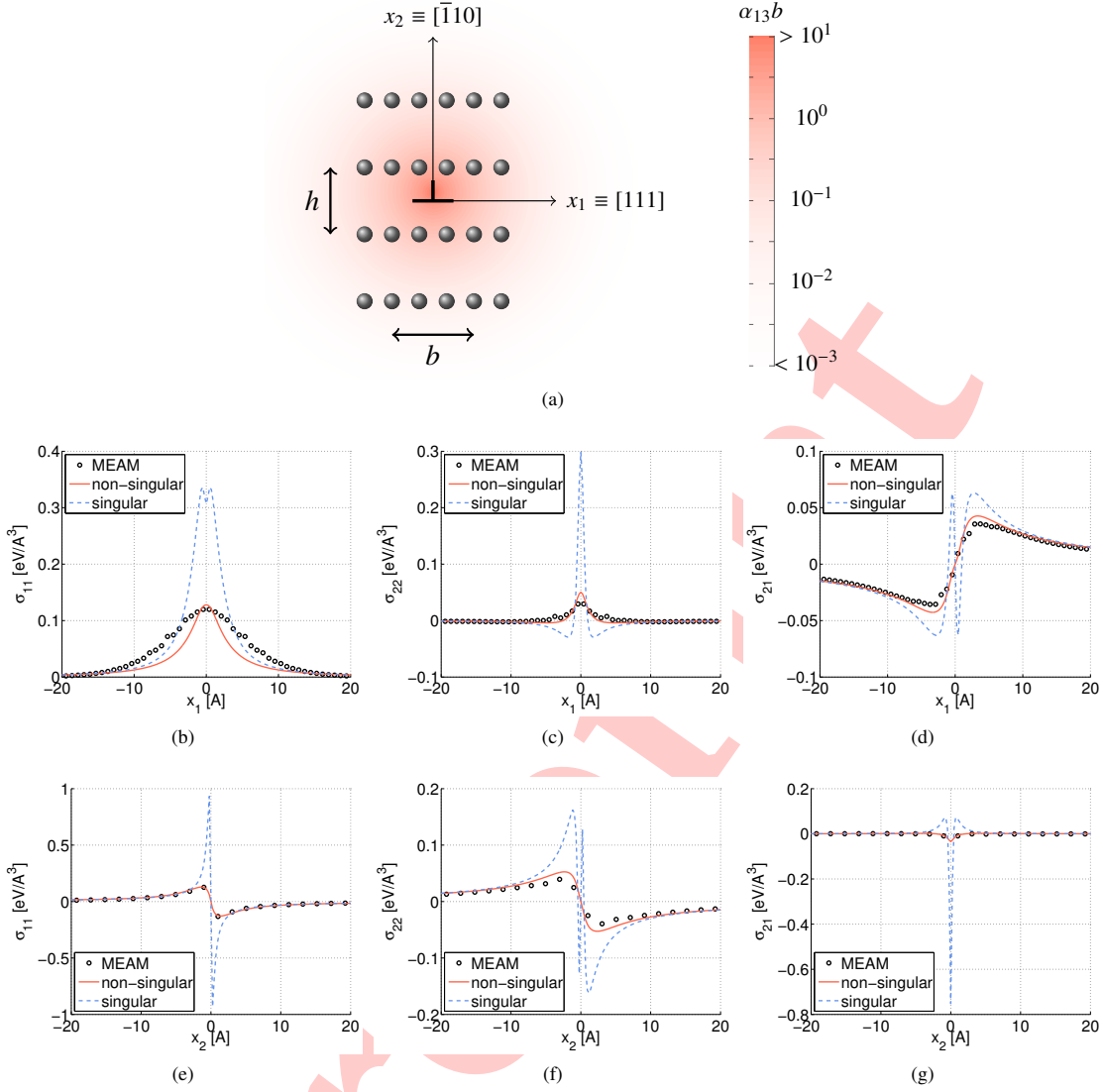

Figure 3: Comparison between the anisotropic singular stress, the anisotropic non-singular stress, and the virial stress of a $1/2[111](\bar{1}10)$ edge dislocation in bcc Fe. Virial stress and elastic constants are obtained from the MEAM interatomic potential of Lee et al. (2001). (a) Nominal position of the edge dislocation used to initially displace the atoms using eq. (58). The background is colored according to Burgers vector density $\alpha_{13}$. (b)-(d) In-plane stress components $\sigma_{11}$, $\sigma_{22}$, $\sigma_{21}$, vs coordinate $x_1$, for $x_2 = -h/2$. (e)-(g) In-plane stress components $\sigma_{11}$, $\sigma_{22}$, $\sigma_{21}$, vs coordinate $x_2$, for $x_1 = -b/6$.

### 4.1. Edge dislocation in bcc Fe

A body-centered cubic (bcc) Fe crystal is constructed using the lattice vectors (expressed in Å in the lattice coordinate system)

$$l^1 = (a, 0, 0) \qquad l^2 = (0, a, 0) \qquad l^3 = (0, 0, a) \tag{60}$$

where $a$ is the lattice parameter describing a unit cell. The unit cell consists of two Fe atoms positioned at $(0, 0, 0)$ and $0.5(l^1 + l^2 + l^3)$, respectively. The lattice is oriented with respect to a fixed global coordinate systems such that the [111] and [$\bar{1}$10] lattice directions are aligned along the global $x_1$ and $x_2$-axes, respectively (see Fig. 3a). A single crystal of size 164.4 Å×134.2 Å×23.3 Å is generated, resulting in a system of 32, 400 atoms. The crystal is relaxed by conjugate gradient minimization of its total energy using the interatomic potential of Lee et al. (2001), a Modified Embedded Atom Method (MEAM) potential. The lattice constant in the relaxed configuration is $a = 2.864$Å. Elastic constants $\mathbb{C}$ and $\mathbb{D}$ are computed in the relaxed configuration. Subsequently, a $1/2[111](\bar{1}10)$ infinite straight edge dislocation with line direction along the $x_3$ axis ([$\bar{1}\bar{1}2$]) is introduced in the system by displacing the reference positions of the atoms by Eq. (58), using an effective Poisson ratio of $\nu = c_{12}/(c_{11} + c_{12}) = 0.36$. The dislocation is located between rows and columns of atoms in the $x_1 x_2$ plane, as illustrated in Fig. 3a. The starting dislocated configuration is again relaxed by conjugate gradient minimization of its total energy. This minimization is performed using the KIM-compliant molecular statics/molecular dynamics code MINIMOL available in the OpenKIM repository of interatomic potentials (Tadmor et al., 2011). Periodic boundary conditions are used in the $x_3$ direction, while in the $x_1 x_2$ plane the outermost layer of atoms is kept fixed. At equilibrium the 0K virial stress is computed according to (59).

The non-singular stress (44c) depends on the tensor $\mathbf{\Lambda}$, which for cubic crystals is isotropic. Using the minimal residual method discussed above, the interatomic potential of Lee et al. (2001) yields the tensor

$$\mathbf{\Lambda} = \begin{bmatrix} 1.756 & 0 & 0 \\ 0 & 1.756 & 0 \\ 0 & 0 & 1.756 \end{bmatrix} \text{Å}^2, \quad (61)$$

which implies a characteristic length $\ell = \sqrt{1.756\text{Å}^2} \approx 1.325$Å, which is about $0.46a$. Because $\mathbf{\Lambda}$ is isotropic, the Burgers vector density in the $x_1 x_2$ plane is also isotropic, as shown in the background of Fig. 3a.

Continuum and atomistic stresses are compared for the row and column of atoms closest to the nominal position of the dislocation. Results for the in-plane stress components are shown in Fig. 3b-3g. From the results, it can be observed that, the anisotropic non-singular fields are in good agreement with the atomistic results, while the classical fields largely overestimate stresses near the dislocation core region. Despite the fact that no fitting procedure was used to obtain these results, the maximum amplitudes of all stress components is captured with reasonable level of accuracy. The most noticeable difference between non-singular and atomistic results is in the breath of the $\sigma_{11}$ profile vs coordinate $x_1$. A striking difference between singular and non-singular solutions is that, for several stress profiles, only the non-singular solution agrees with the sign of the atomistic stress in the core region, as seen in Fig. 3c,3f, and 3g.

### 4.2. Prismatic edge dislocation in hcp Mg

A Mg hcp single crystal is constructed using the lattice vectors

$$l^1 = (a, 0, 0) \qquad l^2 = (-a/2, a\sqrt{3}/2, 0) \qquad l^3 = (0, 0, c). \quad (62)$$

The unit cell consists of two atoms positioned at $(0, 0, 0)$ and $2/3 l^1 + 1/3 l^2 + 1/2 l^3$, respectively. The lattice is oriented with respect to a fixed global coordinate systems such that the [$2\bar{1}\bar{1}0$] and [$01\bar{1}0$] lattice directions are aligned along the global $x_1$ and $x_2$ axes respectively (see Fig. 4a). A single crystal of size 112.3 Å×181.9 Å×33.2 Å is generated resulting in a system of 29, 400 atoms. The crystal is relaxed by conjugate gradient minimization of its total energy using the MEAM interatomic potential of Lee et al. (2001). The lattice constants in the relaxed configuration are $a = 3.209$Å, and $c = 5.197$Å. Elastic constants $\mathbb{C}$ and $\mathbb{D}$ are computed in the relaxed configuration. A $1/3[2\bar{1}\bar{1}0](01\bar{1}0)$ edge dislocation ($a$-type) is created in the system by displacing atoms according to equation (58), with $\nu = 0.27$, as shown in Fig. 4a. The system is allowed to relax to its minimum energy state using conjugate gradient minimization under the same boundary conditions used for bcc Fe.

In the global orientation of Fig. 4, the tensor $\mathbf{\Lambda}$ computed by the minimum residual method reads:

$$\mathbf{\Lambda} = \begin{bmatrix} 1.487 & 0 & 0 \\ 0 & 1.487 & 0 \\ 0 & 0 & 2.760 \end{bmatrix} \text{Å}^2. \quad (63)$$

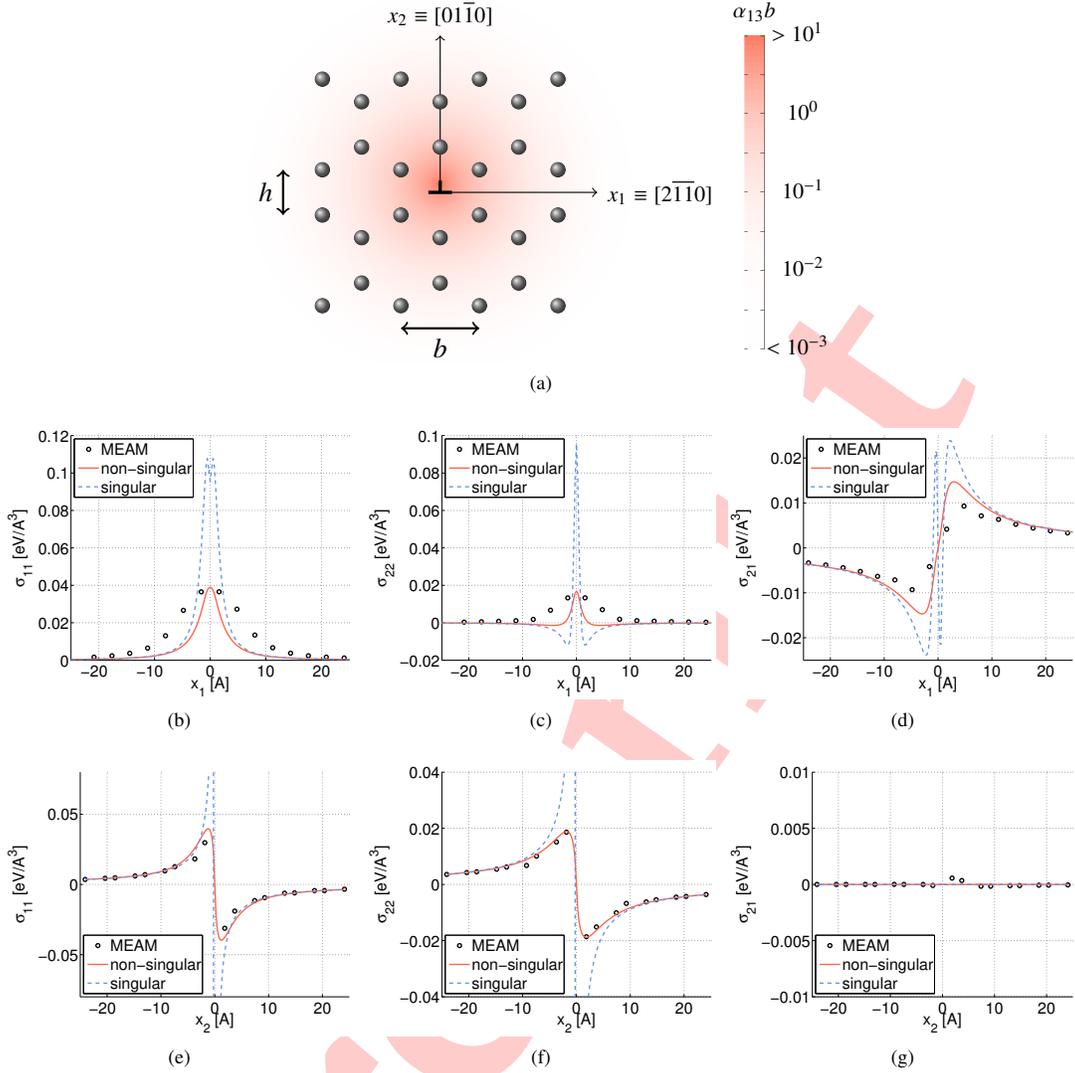

Figure 4: Comparison between the anisotropic singular stress, the anisotropic non-singular stress, and the virial stress of a prismatic $1/3[2\bar{1}\bar{1}0](01\bar{1}0)$ edge dislocation in hcp Mg. Virial stress and elastic constants are obtained from the MEAM interatomic potential of Lee et al. (2001). (a) Nominal position of the edge dislocation used to initially displace the atoms using eq. (58). The background is colored according to Burgers vector density $\alpha_{13}$. (b)-(d) In-plane stress components $\sigma_{11}$, $\sigma_{22}$, $\sigma_{21}$, vs coordinate $x_1$, for $x_2 = -h/2$. (e)-(g) In-plane stress components $\sigma_{11}$, $\sigma_{22}$, $\sigma_{21}$, vs coordinate $x_2$, for $x_1 = 0$.

Note that the two components of $\Lambda$ in the basal plane are identical ($\Lambda_{11} = \Lambda_{22}$), and they correspond to a length scale parameter $\ell_{11} = \sqrt{1.487\text{Å}^2} \approx 1.219\text{Å}$, which is approximately $0.38a$. In the prismatic direction, the length scale parameter is found to be $\ell_{33} = \sqrt{2.760\text{Å}^2} \approx 1.6613\text{Å}$, which is approximately $0.31c$. Because $\Lambda_{11} = \Lambda_{22}$, the Burgers vector density in the $x_1 x_2$ plane is isotropic, as shown in the background of Fig. 4a.

Continuum and atomistic stresses are compared for the row and column of atoms closest to the nominal position of the dislocation. Results for the in-plane stress components are shown in Fig. 4b-4g. Similar to the case of the edge dislocation in bcc Fe, several observations can be made. First, the non-singular stress is in good agreement with the atomistic results for all components of stress, although the profile of $\sigma_{11}$ and $\sigma_{22}$ vs $x_1$ is broader for the atomistic calculations. Second, in contrast to the anisotropic singular solution, the sign of the non-singular solution is always consistent with the atomistic results. Note that the singularity of the classical solution is evident in Fig. 4e-4g since these plots are obtained for a vertical line passing through the origin.

## 4.3. Basal edge dislocation in hcp Mg

Using the same material parameters of the previous section, we now consider a basal edge dislocation in Mg. The displacement field (58) of the full edge dislocation is used to create the dislocation in the atomistic system. During the subsequent relaxation process, the full dislocation dissociates into Shockley partials on the basal plane, according to the reaction

$$\frac{a}{3}[11\bar{2}0] \to \frac{a}{3}[10\bar{1}0] + \frac{a}{3}[01\bar{1}0], \tag{64}$$

and with a stacking fault width of about $8.5a$. A sketch of the corresponding configuration is shown in Fig. 5.

Both singular and non-singular continuum stress fields are obtained summing the stress fields of the two partials, and they are compared to the atomistic results. All in-plane stress components are shown vs global coordinate $x_1$ ([$11\bar{2}0$]), for a value of $x_2$ corresponding to the row of atoms immediately above the glide plane (Fig. 5c-(e)). A slice of the same stress components is also shown vs global coordinate $x_2$ ([0001]), for a value of $x_1$ corresponding to the column of atoms immediately right of the right Shockley partial. Similar to the previous sections, it can be concluded that the non-singular anisotropic results are in good agreement with the atomistic stress. In constrast to the classical stress, the non-singular stress never largely overestimates the atomistic results, and maintains the correct sign in the core region (Fig. 5d, 5g, and 5h).

In order to represent the spread core of each partial, the background of Fig. 5a and 5b is colored according to the components $\alpha_{13}$ and $\alpha_{33}$ of the dislocation density tensor, respectively. In contrast with Figs. 3 and 5, note that the density plots are not isotropic, since they stretched in the $x_2$ direction more then in the $x_1$ direction. This is because, in the global orientation of Fig. 5, the tensor $\Lambda$ has different components $\Lambda_{11}$ and $\Lambda_{22}$:

$$\Lambda = \begin{bmatrix} 1.487 & 0 & 0 \\ 0 & 2.760 & 0 \\ 0 & 0 & 1.487 \end{bmatrix} \text{Å}^2. \tag{65}$$

## 5. Self energy of dislocation loops

The elastic theory of dislocations is sometimes used to compute the elastic component of the free energy of activation in important dislocation processes such as cross-slip (e.g. Ramirez et al., 2012) and nucleation (e.g. Beltz and Freund, 1993). These processes typically involve embryonic dislocation loops whose size may be only a few times the characteristic size of their core. Classical calculations suffer from one of two limitations, either the use of non-singular isotropic solutions, or the use of anisotropic singular solutions based on cut-off radii or equivalent approximations. In contrast, the present theory can be applied directly to compute the non-singular self-energy of dislocation loops in anisotropic materials. The self-energy of a loop is half the expression given in (44d) when loops A and B coincide, that is:

$$W_{\text{self}} = \frac{1}{2} \oiint_{\mathcal{L}} b_j b_i \epsilon_{tkl} \mathbb{C}_{ilmn} \epsilon_{npq} \mathbb{C}_{jprs} F_{skmr}(\boldsymbol{R}) \, dL_q \, dL_t. \tag{66}$$

As a sample calculation, we consider the self-energy of circular $1/2\langle 111\rangle\{110\}$ dislocation loops in bcc Fe. Elastic constants and corresponding length scales have been calculated from the bond-order potential developed by Müller et al. (2007). This potential yields a material with cubic anisotropic ratio $z = 1 - (c_{11} - c_{12})/2c_{44} \approx 0.676$, and a tensor of gradient length scales $\Lambda = \text{diag}\{1.442\text{Å}^2\}$. Energy calculation results are shown in Fig. 6a. Non-singular anisotropic results are compared to non-singular isotropic results by defining a fictitious isotropic material with Lamé parameters $\mu = c_{44} = 0.801\text{eV}/\text{Å}^3$, and $\lambda = c_{12} = 0.886\text{eV}/\text{Å}^3$. Using the same isotropic parameters, we also consider the classical expression for the self-energy of a circular loop given by Hirth and Lothe (1992) in their equation (6-51). In the classical expression, the average energy per unit length scales linearly as $\ln R/b$, where $R$ is the loop radius and $b$ is the magnitude of the Burgers vector, but the line offset is affected significantly by the choice of the "core size" $\rho$.

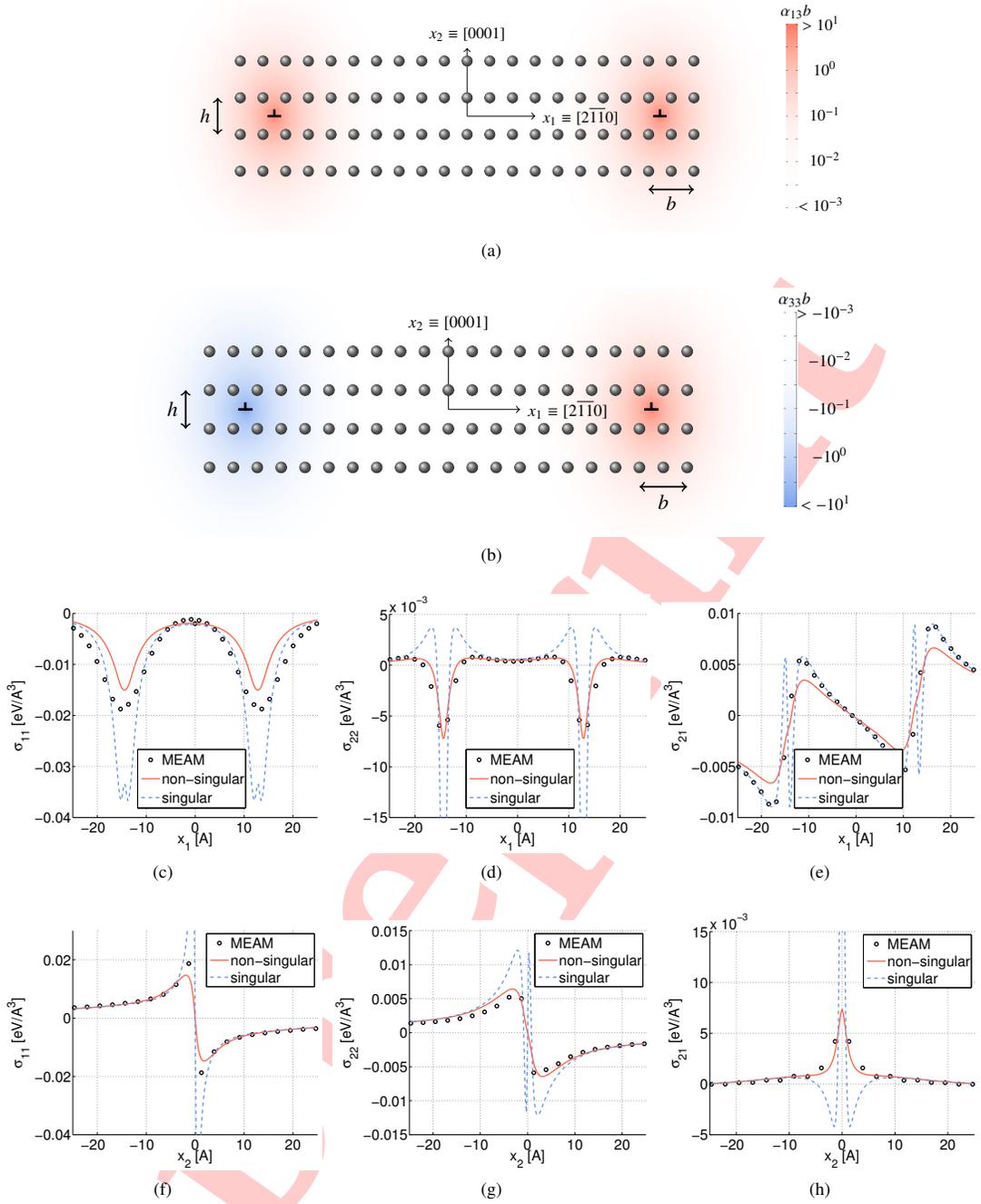

Figure 5: Comparison between the anisotropic singular stress, the anisotropic non-singular stress, and the virial stress of a basal $1/3[2\bar{1}\bar{1}0](0001)$ edge dislocation in hcp Mg. Virial stress and elastic constants are obtained from the MEAM interatomic potential of Lee et al. (2001). (a) Nominal position of Shockley partials used to initially displace the atoms using eq. (58) for each partial. The background is colored according to Burgers vector density $\alpha_{13}$. (b) The background is colored according to Burgers vector density $\alpha_{33}$. (c)-(e) In-plane stress components $\sigma_{11}$, $\sigma_{22}$, $\sigma_{21}$, vs coordinate $x_1$, for $x_2 = h/2$. (f)-(h) In-plane stress components $\sigma_{11}$, $\sigma_{22}$, $\sigma_{21}$, vs coordinate $x_2$, for $x_1 = 4.5b$.

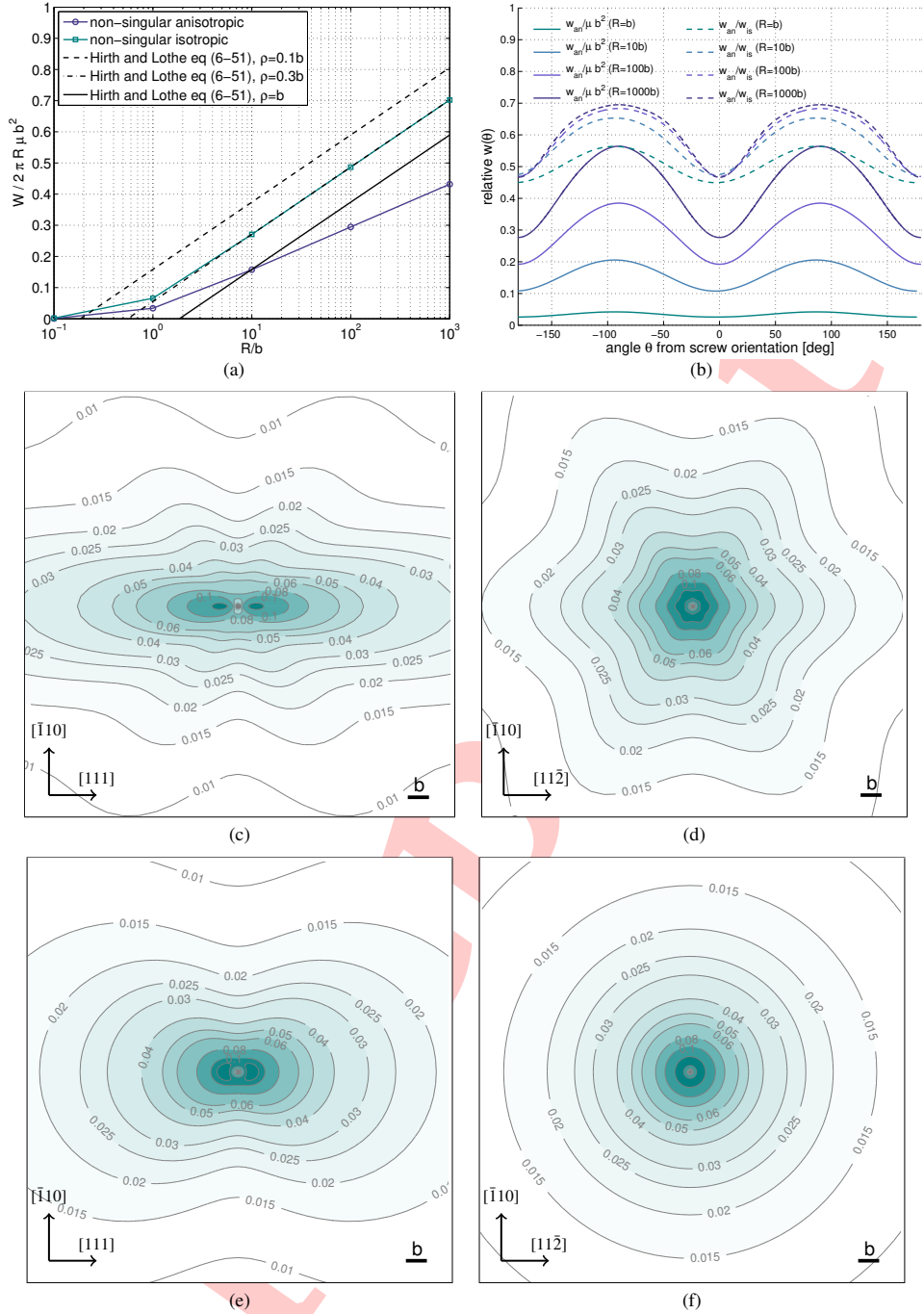

Figure 6: Self energy of circular $1/2\langle 111\rangle\{110\}$ dislocation loops of radius $R$ in bcc Fe. Elastic constants are calculated from the bond-order potential of Müller et al. (2007) (a) Average elastic energy per unit length computed from (66) vs. loop radius. (b) Local elastic energy per unit length computed from (67) vs. angle from the screw orientation and for various loop radii. Norm of the non-singular strain field for the following types of dislocations: (c) anisotropic edge, (d) anisotropic screw, (e) isotropic edge, and (f) isotropic screw.

Moreover the classical expression yields unphysical negative self energy for small loops. The non-singular isotropic solution yields the same scaling factor for large loops, but in this case the energy vanishes as the loop size tends to zero. For the best fit of $\rho$ ($\approx 0.3b$), non-singular and classical solutions start to deviate for $R < 10b$. On the other hand, the non-singular anisotropic calculation results in a scaling factor which is almost half compared to the non-singular isotropic case, and clearly it maintains an analogous trend as the loop size shrinks to zero radius.

From (66), it is also possible to extract the self-energy per unit line of a dislocation, a quantity which depends on the local dislocation character. The self-energy per unit line $w$ can be identified with the kernel of the outermost integral in (66), that is

$$w_{\text{self}}(\boldsymbol{x}) = \frac{1}{2} \oint_{\mathcal{L}} b_j b_i \epsilon_{tkl} \mathbb{C}_{ilmn} \epsilon_{npq} \mathbb{C}_{jprs} F_{skmr}(\boldsymbol{R}) \, \xi'_q(\boldsymbol{x}') \xi_t(\boldsymbol{x}) \, \mathrm{d}L' \,, \tag{67}$$

where $\boldsymbol{\xi}$ is the unit tangent vector to the dislocation line, and as usual $\boldsymbol{R} = \boldsymbol{x} - \boldsymbol{x}'$. Fig. 6b shows the self-energy per unit line as a function of the angle $\theta$ between local tangent vector and the Burgers vector. As expected from classical orientation-dependent line tension models (e.g. Hirth and Lothe, 1992, eq. 6-18), higher energy content is contained in edge-like line elements compared to screw-like line elements. Compared to the isotropic solution, however, this feature is more pronounced in the anisotropic case, as shown by the ratio $w_{\text{an}}/w_{\text{is}}$ of the anisotropic and isotropic energy densities per unit length as a function of the angle $\theta$. Note that the ratio $w_{\text{an}}/w_{\text{is}}$ increases with the size of the loop for the edge orientation, while it remains roughly constant for the screw orientation. For larger loops, screw and edge line elements are well-separated and the local dislocation character and crystallographic orientation determine uniquely the local elastic fields. For example, Figs. 6c and 6d show the norm of the non-singular strain field in the cross section of an edge and screw dislocations, respectively. In both cases, the maximum strain is in the order of 10%, but the strain pattern is significantly different. For the edge dislocation, the strain field stretches in the slip direction, while for the screw dislocation it conforms to the three-fold symmetry of the {110} triad about the ⟨111⟩ slip direction. Clearly the isotropic solution is insensitive to any crystallographic information, as shown for comparison in Figs. 6c and 6d.

## 6. Summary and Conclusion

In this paper we have developed a non-singular theory of three-dimensional dislocation loops valid for anisotropic crystals. The theory is derived in a particular framework of Mindlin's linearized strain-gradient elasticity originally proposed by Lazar and Po (2015b), which assumes the decomposition of the tensor of strain-gradient elastic moduli in the product of the classical tensor of elastic moduli, and a second order symmetric tensor of length-scale parameters ($\mathbb{D} \approx \mathbb{C} \otimes \boldsymbol{\Lambda}$). We showed that, compared to the original Mindlin theory, this assumption tends to preserve the energetic content of uncoupled strain-gradient deformation modes, to the detriment of coupled deformation modes. The attractiveness of the framework is that it constitutes a quasi-classical settings for eigendistortion problems, with the advantage over classical elasticity that the anisotropic Helmholtz-Navier differential operator governing the theory admits a Green's tensor which is intrinsically non-singular (Lazar and Po, 2015b).

The framework accommodates the proposed non-singular anisotropic dislocation theory as a special type of eigendistortion. Similar to existing non-singular theories valid in the isotropic case (Cai et al., 2006; Lazar, 2012, 2013; Po et al., 2014), within our theory dislocations possess a "distributed core" as a consequence of their eigendistortion being obtained from the classical one by convolution with a spreading function. Based on the ansatz that the volume affected by the plastic distortion is determined by the characteristic length scale parameters of the elastic continuum, the spreading function is chosen to be the Green's function of the anisotropic Helmholtz operator. With this choice, all equations of dislocation theory become formally identical to the classical ones, in the sense that they can be obtained from their classical counterparts by replacing the classical (singular) Green's tensor of the Navier operator, $\boldsymbol{G}^0$, with the (non-singular) Green's tensor of the Helmholtz-Navier operator $\boldsymbol{G}$. In real space, the regularized Green tensor carries at a higher computational cost, since it requires an additional integral over the polar angle of the unit sphere. In practical applications such as discrete dislocation dynamics simulations, however, the increased computational cost can be limited to near-core elastic interactions, because the non-singular tensor converges to the classical one a few characteristic lengths away from the origin.

A feature of the proposed dislocation theory is that it entails different core regularization functions for different classes of material symmetry, with six independent length scales for triclinic, four for monoclinic, three for orthorhombic, two for tetragonal, hexagonal, and trigonal, and one for cubic crystals. This is a consequence of the fact that, in order to satisfy material symmetry constraints, the tensor $\boldsymbol{\Lambda}$ has different representations for these crystal classes. Depending on the relative values of the components of $\boldsymbol{\Lambda}$, the Burgers vector density extends differently along different directions, and in general conforms to the lattice of the corresponding crystal class, with diagonal components of $\boldsymbol{\Lambda}$ determining its elongation along the three reference axes, and the off-diagonal components determining its tilting.

As the main result of the paper, the anisotropic non-singular versions of all classical dislocation equations are derived, and in particular: both Volterra and Burgers equations for displacement, the Mura-Willis distortion equation, the Peach-Koehler stress and force equations, and the Blin's formula for the interaction energy between two dislocation loops. All equations are expressed in terms of non-singular line integrals, suitable for numerical implementation. It is important to remark that these anisotropic non-singular solutions tend to their classical counterparts a few characteristic lengths aways from the dislocation core. Therefore their use should be limited in the proximity of the core, while the classical solutions should be used away from the core to increase numerical efficiency.

Because the rank-six tensor of strain-gradient moduli $\mathbb{D}$ and the rank-four tensor of elastic moduli $\mathbb{C}$ can be self-consistently computed from their atomistic representation Admal et al. (2017), two methods were proposed to estimate the tensor $\boldsymbol{\Lambda}$, namely a minimal residual method, and a projection method. The two methods yield comparable values for the components of $\boldsymbol{\Lambda}$ for cubic and hexagonal crystals considered in this paper. Molecular statics calculations of the stress field in both cubic and hexagonal crystals were compared to the proposed anisotropic non-singular dislocation theory, without using any fitting procedure. It was shown that the anisotropic non-singular fields are in good agreement with the atomistic results, while the classical fields largely overestimate stresses near the dislocation core region. Moreover, for several stress components, only the non-singular solution agrees with the sign of the atomistic stress in the core region. In order to demonstrate the feasibility of the non-singular equations in DDD applications, all elastic calculations were performed using the DDD method, that is by numerical integration of the stress kernel along the dislocation lines. Finally, we computed the self energy of circular dislocation loops in bcc Fe for loop sizes ranging over five orders of magnitude. For larger loops, the energy per unit length scales classically as $\ln R/b$, but with significantly different proportionality constants for the isotropic and anisotropic solutions. Compared to the non-singular solution, the best fit for the classical expression breaks down when the size of the loop is in the order of ten times the characteristic length scale of the material. The line energy per unit length as a function of the local dislocation character was also evaluate numerically, and it was shown that, compared to isotropic calculations, anisotropy increases the difference between screw and edge line elements.

## Acknowledgements

Giacomo Po and Nasr Ghoniem acknowledge the support of the U.S. Department of Energy, Office of Fusion Energy, through the DOE award number DE-FG02-03ER54708 at UCLA, and the Air Force Office of Scientific Research (AFOSR), through award number FA9550-11-1-0282 with UCLA. Giacomo Po acknowledges the support of the National Science Foundation, Division of Civil, Mechanical and Manufacturing Innovation (CMMI), through award number 1563427 with UCLA. Markus Lazar gratefully acknowledges the grants obtained from the Deutsche Forschungsgemeinschaft (Grant Nos. La1974/2-2, La1974/3-1, La1974/4-1).

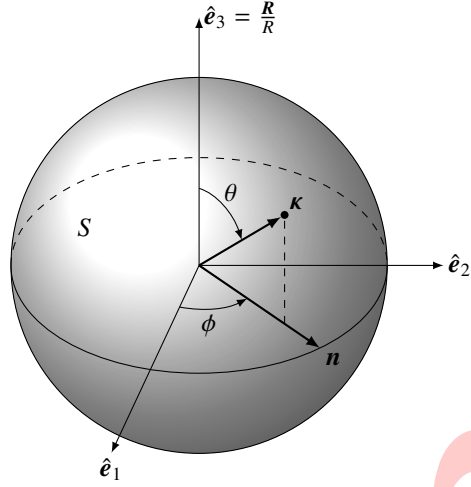

Figure A.7: The unit sphere in Fourier space. The unit vector $\boldsymbol{\kappa}(\theta, \phi)$ is defined by the azimuth angle $\phi$, and the zenith angle $\theta$ measured from the axis $\hat{\boldsymbol{e}}_3 = \boldsymbol{R}/R$.

## Appendix A. The Green's tensor and the F-tensor in classical anisotropic elasticity

In classical elasticity, the Green's tensor of the anisotropic Navier operator $G_{ij}^0$ satisfies the following inhomogeneous PDE:

$$L_{ik} G_{kj}^0 + \delta_{ij}\delta = 0\,. \tag{A.1}$$

In Fourier space[9] this reads:

$$\hat{G}^0_{ik}(\boldsymbol{k}) = \frac{1}{k^2}\,\hat{L}^{-1}_{ik}(\boldsymbol{\kappa})\,. \tag{A.3}$$

where $\hat{L}_{ik}(\boldsymbol{\kappa}) = \mathbb{C}_{ijkl}\kappa_j\kappa_l$, $\boldsymbol{\kappa} = \boldsymbol{k}/k$, and $k = \sqrt{\boldsymbol{k}\cdot\boldsymbol{k}}$. The Green's tensor in real space is obtained by inverse Fourier transform. Expressing the elementary volume element in Fourier space as $\mathrm{d}\hat{V} = k^2\,\mathrm{d}k\,\mathrm{d}\omega$, where $\mathrm{d}\omega$ is an elementary surface element of the unit sphere $\mathcal{S}$, we obtain:

$$G^0_{ik}(\boldsymbol{R}) = \frac{1}{(2\pi)^3}\int_{\mathbb{R}^3}\frac{\hat{L}^{-1}_{ik}(\boldsymbol{\kappa})}{k^2}\,\mathrm{e}^{\mathrm{i}\boldsymbol{k}\cdot\boldsymbol{x}}\,\mathrm{d}\hat{V} = \frac{1}{(2\pi)^3}\int_{\mathcal{S}}\hat{L}^{-1}_{kl}(\boldsymbol{\kappa})\int_0^\infty\cos(k\boldsymbol{\kappa}\cdot\boldsymbol{R})\,\mathrm{d}k\,\mathrm{d}\omega = \frac{1}{8\pi^2 R}\int_{\mathcal{S}}\hat{L}^{-1}_{kl}(\boldsymbol{\kappa})\,\delta(\boldsymbol{\kappa}\cdot\boldsymbol{R})\,\mathrm{d}\omega\,.$$

Choosing a reference system with $\hat{\boldsymbol{e}}_3$ aligned with $\boldsymbol{R}$, as shown in Fig. A.7, and using the sifting property of the Dirac $\delta$-function, we finally obtain the expression for the Green tensor as:

$$G^0_{ik}(\boldsymbol{R}) = \frac{1}{8\pi^2 R}\int_0^{2\pi}\hat{L}^{-1}_{ik}(\boldsymbol{n})\,\mathrm{d}\phi\,. \tag{A.4}$$

Here, $\boldsymbol{n}$ indicates a unit vector on the equatorial plane of the unit sphere in Fourier space. This result was first obtained by Lifshitz and Rosenzweig (1947) and Synge (1957).

The classical $\boldsymbol{F}$-tensor, introduced by Kirchner (1984), is defined by Eq. (15), which in Fourier space reads:

$$\hat{F}^0_{ijkl} = -\hat{G}^0_{kl}k_ik_j\hat{G}^\Delta = -\frac{1}{k^2}\,\hat{L}^{-1}_{kl}(\boldsymbol{\kappa})\,k_ik_j\,\frac{1}{k^2} = -\frac{1}{k^2}\hat{L}^{-1}_{kl}(\boldsymbol{\kappa})\,\kappa_i\kappa_j\,. \tag{A.5}$$

The classical $\boldsymbol{F}$-tensor in real space is obtained by inverse Fourier transform:

$$\begin{aligned}F^0_{ijkl}(\boldsymbol{R}) &= -\frac{1}{(2\pi)^3}\int_{\mathbb{R}^3}\frac{1}{k^2}\,\hat{L}^{-1}_{kl}(\boldsymbol{\kappa})\,\kappa_i\kappa_j\,\mathrm{e}^{\mathrm{i}\boldsymbol{k}\cdot\boldsymbol{R}}\,\mathrm{d}\hat{V} = -\frac{1}{(2\pi)^3}\int_{\mathcal{S}}\hat{L}^{-1}_{kl}(\boldsymbol{\kappa})\,\kappa_i\kappa_j\int_0^\infty\cos(k\boldsymbol{\kappa}\cdot\boldsymbol{R})\,\mathrm{d}k\,\mathrm{d}\omega\\ &= -\frac{1}{8\pi^2 R}\int_{\mathcal{S}}\hat{L}^{-1}_{kl}(\boldsymbol{\kappa})\,\kappa_i\kappa_j\,\delta(\boldsymbol{\kappa}\cdot\boldsymbol{R})\,\mathrm{d}\omega\,.\end{aligned} \tag{A.6}$$

In the reference system of Fig. A.7, we finally obtain:

$$F^0_{ijkl}(\boldsymbol{R}) = -\frac{1}{8\pi^2 R}\int_0^{2\pi}\hat{L}^{-1}_{kl}(\boldsymbol{n})\,n_in_j\,\mathrm{d}\phi\,. \tag{A.7}$$

### Appendix B. The Green's tensor and the F-tensor in gradient anisotropic elasticity

In gradient elasticity with separable weak non-locality, the Green's tensor of the anisotropic Helmholtz-Navier operator $G_{ij}$ satisfies the following inhomogeneous PDE:

$$L_{ik}L G_{kj}(\boldsymbol{R}) + \delta_{ij}\delta(\boldsymbol{R}) = 0\,, \tag{B.1}$$

which in Fourier space reads

$$\hat{G}_{kl}(\boldsymbol{k}) = \frac{1}{1+k^2\Lambda_{mn}\kappa_m\kappa_n}\,\frac{\hat{L}^{-1}_{kl}(\boldsymbol{\kappa})}{k^2} = \frac{1}{1+k^2\lambda^2(\boldsymbol{\kappa})}\,\frac{\hat{L}^{-1}_{kl}(\boldsymbol{\kappa})}{k^2}\,, \tag{B.2}$$

---

[9] The Fourier transform and its inverse are defined as, respectively (Vladimirov, 1971):

$$\hat{f}(\boldsymbol{k}) = \int_{\mathbb{R}^3}f(\boldsymbol{x})\,\mathrm{e}^{-\mathrm{i}\boldsymbol{k}\cdot\boldsymbol{x}}\,\mathrm{d}V\,, \qquad\qquad f(\boldsymbol{x}) = \frac{1}{(2\pi)^3}\int_{\mathbb{R}^3}\hat{f}(\boldsymbol{k})\,\mathrm{e}^{\mathrm{i}\boldsymbol{k}\cdot\boldsymbol{x}}\,\mathrm{d}\hat{V}\,. \tag{A.2}$$

where $\lambda^2(\boldsymbol{\kappa}) = \Lambda_{mn}\kappa_m\kappa_n$. The Green's tensor in real space is obtained by inverse Fourier transform (Lazar and Po, 2015b):

$$\begin{aligned} G_{ij}(\boldsymbol{R}) &= \frac{1}{(2\pi)^3} \int_{\mathbb{R}^3} \frac{1}{k^2(1+k^2\lambda^2(\boldsymbol{\kappa}))} \hat{L}_{kl}^{-1}(\boldsymbol{\kappa}) e^{i\boldsymbol{k}\cdot\boldsymbol{R}} \, d\hat{V} \\ &= \frac{1}{(2\pi)^3} \int_{\mathcal{S}} \hat{L}_{kl}^{-1}(\boldsymbol{\kappa}) \int_0^\infty \frac{\cos(k\boldsymbol{\kappa}\cdot\boldsymbol{R})}{1+\lambda^2(\boldsymbol{\kappa})k^2} \, dk \, d\omega = \frac{1}{16\pi^2} \int_{\mathcal{S}} \hat{L}_{kl}^{-1}(\boldsymbol{\kappa}) \frac{e^{-|\boldsymbol{\kappa}\cdot\boldsymbol{R}|/\lambda(\boldsymbol{\kappa})}}{\lambda(\boldsymbol{\kappa})} \, d\omega \, . \end{aligned} \quad \text{(B.3)}$$

In the reference system of Fig. A.7, letting $q = \cos\theta$ we have:

$$G_{ij}(\boldsymbol{R}) = \frac{1}{8\pi^2} \int_0^{2\pi} \int_0^1 \hat{L}_{ij}^{-1}(\boldsymbol{\kappa}) \frac{e^{-Rq/\lambda(\boldsymbol{\kappa})}}{\lambda(\boldsymbol{\kappa})} \, dq \, d\phi \, . \quad \text{(B.4)}$$

The gradient of the Green tensor is simply:

$$G_{ij,m}(\boldsymbol{R}) = -\frac{1}{8\pi^2} \int_0^{2\pi} \int_0^1 \hat{L}_{ij}^{-1}(\boldsymbol{\kappa}) \kappa_m \frac{e^{-Rq/\lambda(\boldsymbol{\kappa})}}{\lambda^2(\boldsymbol{\kappa})} \, dq \, d\phi \, . \quad \text{(B.5)}$$

The $\boldsymbol{F}$ tensor in gradient elasticity is defined by Eq. (40) which In Fourier space reads:

$$\hat{F}_{ijkl} = -\hat{G}_{kl} k_i k_j \hat{G}^\Delta = -\hat{G}_{kl}^0 \hat{G}^L k_i k_j \hat{G}^\Delta = -\frac{1}{k^2} \frac{1}{1+k^2\lambda^2(\boldsymbol{\kappa})} \hat{L}_{kl}^{-1}(\boldsymbol{\kappa}) k_i k_j \frac{1}{k^2} = -\frac{1}{k^2(1+k^2\lambda^2(\boldsymbol{\kappa}))} \hat{L}_{kl}^{-1}(\boldsymbol{\kappa}) \kappa_i \kappa_j \, . \quad \text{(B.6)}$$

The $\boldsymbol{F}$ tensor in real space is obtained by inverse Fourier transform:

$$\begin{aligned} F_{ijkl}(\boldsymbol{R}) &= -\frac{1}{(2\pi)^3} \int_{\mathbb{R}^3} \frac{1}{k^2(1+k^2\lambda^2(\boldsymbol{\kappa}))} \hat{L}_{kl}^{-1}(\boldsymbol{\kappa}) \kappa_i \kappa_j \, e^{i\boldsymbol{k}\cdot\boldsymbol{R}} \, d\hat{V} \\ &= -\frac{1}{(2\pi)^3} \int_{\mathcal{S}} \hat{L}_{kl}^{-1}(\boldsymbol{\kappa}) \kappa_i \kappa_j \int_0^\infty \frac{\cos(k\boldsymbol{\kappa}\cdot\boldsymbol{R})}{1+\lambda^2(\boldsymbol{\kappa})k^2} \, dk \, d\omega = -\frac{1}{16\pi^2} \int_{\mathcal{S}} \hat{L}_{kl}^{-1}(\boldsymbol{\kappa}) \kappa_i \kappa_j \frac{e^{-|\boldsymbol{\kappa}\cdot\boldsymbol{R}|/\lambda(\boldsymbol{\kappa})}}{\lambda(\boldsymbol{\kappa})} \, d\omega \, . \end{aligned} \quad \text{(B.7)}$$

In the reference system of Fig. A.7, we finally obtain:

$$F_{ijkl}(\boldsymbol{R}) = -\frac{1}{8\pi^2} \int_0^{2\pi} \int_0^1 \hat{L}_{kl}^{-1}(\boldsymbol{\kappa}) \kappa_i \kappa_j \frac{e^{-Rq/\lambda(\boldsymbol{\kappa})}}{\lambda(\boldsymbol{\kappa})} \, dq \, d\phi \, . \quad \text{(B.8)}$$

### Appendix C. The Green's function of the anisotropic Laplace-Helmholtz equation: $G^{\Delta L} = G^\Delta * G^L$

The Green's function of the one-fold anisotropic Laplace-Helmholtz operator $\Delta L$, which means the isotropic Laplace operator $\Delta$ and the anisotropic Helmholtz operator $L$, is defined by

$$\Delta L G^{\Delta L}(\boldsymbol{R}) = \delta(\boldsymbol{R}) \, . \quad \text{(C.1)}$$

In Fourier space, the Green's function reads

$$\hat{G}^{\Delta L} = \hat{G}^L \hat{G}^\Delta = \frac{1}{1+\Lambda_{mn}k_m k_n} \frac{1}{k^2} \, . \quad \text{(C.2)}$$

In real space the Green's function of the operator $\Delta L$ is obtained as

$$G^{\Delta L}(\boldsymbol{R}) = -\frac{1}{8\pi^2} \int_0^{2\pi} \int_0^1 \frac{e^{-Rq/\lambda(\boldsymbol{\kappa})}}{\lambda(\boldsymbol{\kappa})} \, dq \, d\phi \, . \quad \text{(C.3)}$$

The gradient of Eq. (C.3) reads

$$G_{,m}^{\Delta L}(\boldsymbol{R}) = \frac{1}{8\pi^2} \int_0^{2\pi} \int_0^1 \kappa_m \frac{e^{-Rq/\lambda(\boldsymbol{\kappa})}}{\lambda^2(\boldsymbol{\kappa})} \, dq \, d\phi \, . \quad \text{(C.4)}$$

## Appendix D. Line integral representation of the solid angle in anisotropic gradient elasticity

Following Lazar and Po (2014), we give here the anisotropic solid angle as line integral. The solid angle with weak anisotropic nonlocality reduces to a line integral of the monopole vector potential $A_k$ and a contribution due to the fictitious vector field $v_i^{(f)}$

$$\Omega(\boldsymbol{x}) = \oint_{\mathcal{L}} A_k(\boldsymbol{x} - \boldsymbol{x}') \, \mathrm{d}L'_k - 4\pi \int_{\mathcal{S}} \int_C G^L(\boldsymbol{x} - \boldsymbol{x}') \, \mathrm{d}L'_i \, \mathrm{d}S_i \,, \tag{D.1}$$

where $G^L$ is the Green's function of the anisotropic Helmholtz equation. $G^L(\boldsymbol{x} - \boldsymbol{x}')$ is non-zero for $\boldsymbol{x} = \boldsymbol{x}'$ and for $\boldsymbol{x}$ different than $\boldsymbol{x}'$ near the Dirac string.

The corresponding monopole vector potential satisfies the following inhomogeneous one-fold anisotropic Laplace-Helmholtz equation (see also Lazar and Po (2014))

$$\Delta L A_k = -\epsilon_{klm} \partial_l v_m^{(f)0} \,, \tag{D.2}$$

where the fictitious singular vector field $v_i^{(f)0}$ is given by

$$v_i^{(f)0}(\boldsymbol{x}) = 4\pi \int_C \delta(\boldsymbol{x} - \boldsymbol{s}) \, \mathrm{d}s_i \equiv 4\pi \, \delta_i(C) \tag{D.3}$$

and $C$ is a curve, called the "Dirac string", starting at $-\infty$ and ending at the origin and $\delta_i(C)$ is the $\delta$-function along the Dirac string. In Eq. (D.2), $L$ denotes the anisotropic Helmholtz operator. The solution of Eq. (D.2) can be written as convolution

$$A_k = -G^{\Delta L} * (\epsilon_{klm} \partial_l v_m^{(f)0}) = -4\pi \epsilon_{klm} \partial_l G^{\Delta L} * \delta_m(C) = -4\pi \epsilon_{klm} \oint_C \partial_l G^{\Delta L}(\boldsymbol{x} - \boldsymbol{s}) \, \mathrm{d}s_m \,. \tag{D.4}$$

The explicit form of the monopole vector field reads

$$A_k(\boldsymbol{x}) = -\frac{1}{2\pi} \, \epsilon_{klm} \oint_C \int_0^{2\pi} \int_0^1 \kappa_l \, \frac{\mathrm{e}^{-|\boldsymbol{x}-\boldsymbol{s}|q/\lambda(\boldsymbol{\kappa})}}{\lambda^2(\boldsymbol{\kappa})} \, \mathrm{d}q \, \mathrm{d}\phi \, \mathrm{d}s_m \,. \tag{D.5}$$

## Appendix E. Anisotropic non-singular theory in classical elasticity

Inspired by the theory developed in the section 2.4, it is also possible to construct a regularized version of classical dislocation theory in anisotropic media. The approach can be regarded as the anisotropic extension of the theory developed by Cai et al. (2006) in the isotropic case. Suppose that we replace the classical dislocation eigendistortion (20) with a new function obtained by convolution with a spreading function $\tilde{h}(\boldsymbol{x})$:

$$\beta_{kl}^P(\boldsymbol{x}) = \tilde{h}(\boldsymbol{x}) * \beta_{kl}^{P,0}(\boldsymbol{x}) \,. \tag{E.1}$$

As a consequence we obtain a dislocation density tensor

$$\alpha_{ij}(\boldsymbol{x}) = \tilde{h}(\boldsymbol{x}) * \alpha_{ij}^0(\boldsymbol{x}) \,, \tag{E.2}$$

where $\alpha_{ij}^0(\boldsymbol{x})$ is the classical term given in (21). Using (E.1) and (E.2) in place of (20) and (21) in the field equations of classical dislocation theory (22) we obtain:

$$\tilde{u}_i(\boldsymbol{x}) = -\frac{b_i \tilde{\Omega}(\boldsymbol{x})}{4\pi} - \oint_{\mathcal{L}} \mathbb{C}_{mnpq} \epsilon_{jqr} b_p \left( F^0_{jnim} * \tilde{h} \right)(\boldsymbol{R}) \, \mathrm{d}L'_r \tag{E.3a}$$

$$\tilde{\beta}_{ij}(\boldsymbol{x}) = \oint_{\mathcal{L}} \mathbb{C}_{mnpq} \epsilon_{jqr} \left( G^0_{im,n} * \tilde{h} \right)(\boldsymbol{R}) b_p \, \mathrm{d}L'_r \tag{E.3b}$$

$$\tilde{\sigma}_{ij}(\boldsymbol{x}) = \oint_{\mathcal{L}} \mathbb{C}_{ijkl} \mathbb{C}_{mnpq} \epsilon_{lqr} \left( G^0_{km,n} * \tilde{h} \right)(\boldsymbol{R}) b_p \, \mathrm{d}L'_r \tag{E.3c}$$

$$\widetilde{\widetilde{W}}_{AB} = \oint_{\mathcal{L}_A} \oint_{\mathcal{L}_B} \epsilon_{jkl} \mathbb{C}_{ilmn} \epsilon_{npq} \mathbb{C}_{rstp} \left( F^0_{skmr} * \tilde{h} * \tilde{h} \right)(\boldsymbol{R}) \, b_t^A b_i^B \, \mathrm{d}L_q^A \, \mathrm{d}L_j^B \tag{E.3d}$$

$$\widetilde{\widetilde{\mathcal{F}}}_k = \oint_{\mathcal{L}} \epsilon_{kjm} \left( \tilde{\sigma}_{ij} * \tilde{h} \right) b_i \, \mathrm{d}L_m = \oint_{\mathcal{L}} \epsilon_{kjm} \left( \sigma^0_{ij} * \tilde{h} * \tilde{h} \right) b_i \, \mathrm{d}L_m \,, \tag{E.3e}$$

where $\boldsymbol{R} = \boldsymbol{x} - \boldsymbol{x}'$ and $\tilde{\Omega}(\boldsymbol{x}) = 4\pi \int_{\mathcal{S}} \left( G^{\Delta}_{,j} * \tilde{h} \right)(\boldsymbol{R}) \, \mathrm{d}A'_j$.

In order to maintain the line integral representation of the dislocation field equations, we now need to choose the spreading function $\tilde{h}$ in a way that the convolution integrals can be carried out analytically. In this perspective, two choices are of interest. On the one hand, if we choose $\tilde{h} = G^L$, then the first three equations in (E.3) become identical to the first three equations in (44). On the other hand, the choice $\tilde{h} * \tilde{h} = G^L$ makes the last two equations in (E.3) identical to the last two equations in (44). Therefore, it is seen that the Green's function $G^L$ can be used to regularize the dislocation fields also in the classical settings, with the drawback that only a certain subset of (E.3) retains its line integral form for a certain choice of the spreading function. This dichotomy does not exist in the gradient-elastic framework discussed in sections 2.3 and 2.4.

## Appendix F. Identities

The fourth-order partial differential equation (29) can be reduced to a system of second-order partial differential equations (see Lazar (2014)). More precisely, Eq. (29) is equivalent to the inhomogeneous Navier equation

$$L_{ik} u^0_k = \mathbb{C}_{ijkl} \beta^{P,0}_{kl,j} \tag{F.1}$$

and the two inhomogeneous Helmholtz equations

$$L u_i = u^0_i, \tag{F.2}$$

$$L \beta^P_{ij} = \beta^{P,0}_{ij}. \tag{F.3}$$

Eq. (F.1) is the classical Navier equation (9). In addition using (F.3), Eq. (29) can be written equivalently into the following system of PDEs

$$L_{ik} L u_k = \mathbb{C}_{ijkl} \beta^{P,0}_{kl,j}, \tag{F.4}$$

$$L \beta^P_{ij} = \beta^{P,0}_{ij}. \tag{F.5}$$

From Eq. (29), a tensorial inhomogeneous Navier-Helmholtz equation can be derived (e.g., Lazar (2013, 2014))

$$L_{ik} L \beta_{km} = -\mathbb{C}_{ijkl} \epsilon_{mlr} L \alpha_{kr,j}. \tag{F.6}$$

The fourth-order partial differential equation (F.6) can also be reduced to a system of second-order partial differential equations (see Lazar (2014)). Thus, Eq. (29) is equivalent to the inhomogeneous Navier

$$L_{ik} \beta^0_{km} = -\mathbb{C}_{ijkl} \epsilon_{mlr} \alpha^0_{kr,j} \tag{F.7}$$

and the two inhomogeneous Helmholtz equations

$$L \beta_{ij} = \beta^0_{ij}, \tag{F.8}$$

$$L \alpha_{ij} = \alpha^0_{ij}. \tag{F.9}$$

Eq. (F.7) is the classical Navier equation (F.1). In addition using Eq. (F.9), Eq. (F.6) can be written equivalently into the following system of PDEs

$$L_{ik} L \beta_{km} = -\mathbb{C}_{ijkl} \epsilon_{mlr} \alpha^0_{kr,j}, \tag{F.10}$$

$$L \alpha_{ij} = \alpha^0_{ij}. \tag{F.11}$$

27